\documentclass[prl,twocolumn,superscriptaddress,secnumarabic,balancelastpage,amsmath,amssymb]{revtex4-2}

\usepackage{amsmath}
\usepackage{amsthm}
\usepackage{braket}
\usepackage{gensymb}
\usepackage{graphicx}
\usepackage{makecell}

\usepackage{verbatim}
\usepackage{graphicx}
\usepackage[colorlinks,linkcolor=black,citecolor=black,urlcolor=black,filecolor=black]{hyperref}
\usepackage{physics}
\usepackage{ulem}

\newcommand{\e}{\mathfrak{e}}
\newcommand{\lket}[1]{|{\overline{#1}}\rangle}
\newcommand{\ooverline}[1]{\overline{\overline{#1}}}

\newcommand{\tcal}[1]{\tilde{\mathcal{#1}}}

\begin{document}

\title{Low-Overhead Transversal Fault Tolerance for Universal Quantum Computation}

\author{Hengyun Zhou}
\thanks{These authors contributed equally}
\email{hyzhou@quera.com}
\affiliation{QuEra Computing Inc., 1284 Soldiers Field Road, Boston, MA, 02135, US}
\affiliation{Department of Physics, Harvard University, Cambridge, Massachusetts 02138, USA}

\author{Chen Zhao}
\thanks{These authors contributed equally}
\affiliation{QuEra Computing Inc., 1284 Soldiers Field Road, Boston, MA, 02135, US}

\author{Madelyn Cain}
\affiliation{Department of Physics, Harvard University, Cambridge, Massachusetts 02138, USA}

\author{Dolev Bluvstein}
\affiliation{Department of Physics, Harvard University, Cambridge, Massachusetts 02138, USA}

\author{Nishad Maskara}
\affiliation{Department of Physics, Harvard University, Cambridge, Massachusetts 02138, USA}

\author{Casey Duckering}
\affiliation{QuEra Computing Inc., 1284 Soldiers Field Road, Boston, MA, 02135, US}

\author{Hong-Ye Hu}
\affiliation{Department of Physics, Harvard University, Cambridge, Massachusetts 02138, USA}

\author{Sheng-Tao Wang}
\affiliation{QuEra Computing Inc., 1284 Soldiers Field Road, Boston, MA, 02135, US}

\author{Aleksander Kubica}
\affiliation{AWS Center for Quantum Computing, Pasadena, California 91125, USA}
\affiliation{California Institute of Technology, Pasadena, California 91125, USA}
\affiliation{Department of Applied Physics, Yale University, New Haven, Connecticut 06511, USA}

\author{Mikhail D. Lukin}
\email{lukin@physics.harvard.edu}
\affiliation{Department of Physics, Harvard University, Cambridge, Massachusetts 02138, USA}

\begin{abstract}
Fast, reliable logical operations are essential for realizing useful quantum computers~\cite{dalzell2023,beverland2022assessing,babbush2021focus}.
By redundantly encoding logical qubits into many physical qubits and using syndrome measurements to detect and correct errors, one can achieve low logical error rates.
However, for many practical quantum error correcting (QEC) codes such as the surface code, due to syndrome measurement errors, standard constructions require multiple extraction rounds---on the order of the code distance $d$---for fault-tolerant computation, particularly considering fault-tolerant state preparation~\cite{gottesman2010introduction,gottesman2013fault,steane1996error,shor1996fault,kitaev2003fault,bravyi1998quantum, dennis2002topological,horsman2012surface,litinski2019game,fowler2012surface,litinski2022active}. 
Here, we show that logical operations can be performed fault-tolerantly with only a constant number of extraction rounds for a broad class of QEC codes, including the surface code with magic state inputs and feed-forward, to achieve ``transversal algorithmic fault tolerance".
Through the combination of transversal operations~\cite{shor1996fault} and novel strategies for correlated decoding~\cite{cain2024correlated}, despite only having access to partial syndrome information, we prove that the deviation from the ideal logical measurement distribution can be made exponentially small in the distance, even if the instantaneous quantum state cannot be made close to a logical codeword due to measurement errors.
We supplement this proof with circuit-level simulations in a range of relevant settings, demonstrating the fault tolerance and competitive performance of our approach.
Our work sheds new light on the theory of quantum fault tolerance and has the potential to reduce the space-time cost of practical fault-tolerant quantum computation by over an order of magnitude.
\end{abstract}
\maketitle


Quantum computers have the potential to solve certain computational problems much faster than their classical counterparts~\cite{dalzell2023,gidney2019how}.
Since most known applications require quantum computers with extremely low error rates, quantum error correction (QEC) and strategies for fault-tolerant quantum computing (FTQC) are necessary.
These methods encode logical quantum information into a QEC code involving many physical qubits, such that the lowest weight logical error has weight equal to the code distance $d$ and is therefore unlikely.

\begin{figure}
\centering
\includegraphics{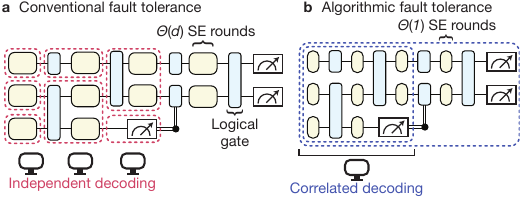}
\caption{\textbf{Transversal algorithmic fault tolerance.}
(a)~Conventional FT analysis separately examines each gadget (red boxes) in the circuit and ensures they are individually FT~\cite{gottesman2010introduction,shor1996fault,aharonov1999fault}.
This typically requires $\Theta(d)$ syndrome extraction (SE) rounds to achieve FT.
(b)~Transversal algorithmic FT directly uses all accessible syndrome information up to a logical measurement (blue box), and guarantees FT of the logical measurement result, even if the gadgets are not individually FT and if future syndrome information is not yet accessible (partial decoding).
We realize this through transversal operations, and only require a single SE round per logical operation.
}
\label{fig:fig1}
\end{figure}

Performing large-scale computation, however, comes with significant overhead~\cite{gidney2019how,beverland2022assessing}.
By performing syndrome extraction (SE), one can reveal error information and use a classical decoder to correct physical errors in software and interpret logical measurement results.
However, in the presence of noisy syndrome measurements~\cite{gottesman2010introduction,gottesman2013fault,steane1996error,shor1996fault,dennis2002topological}, one typically requires a number of SE rounds that scales linearly in $d$, i.e., $\Theta(d)$~\footnote{The notation $g(x)=\Theta(f(x))$ indicates that two functions $f(x)$ and $g(x)$ have the same asymptotic scaling with $x$, or more precisely, that there exists some constants $c_1$ and $c_2$ such that $c_1f(x)\leq g(x)\leq c_2f(x)$ for sufficiently large $x$.} (see Fig.~\ref{fig:fig1}(a)).
This is the case, for example, for the celebrated surface code~\cite{kitaev2003fault,bravyi1998quantum,dennis2002topological}, one of the leading candidates for practical FTQC due to its simple 2D layout and competitive error thresholds.
Here, even the most basic task of fault-tolerantly preparing a logical $\lket{0}$ state requires $\Theta(d)$ SE rounds~\cite{hastings2011topological},
reducing the logical clock speed by a factor proportional to the code distance (typically 10\,--100~\cite{gidney2019how,litinski2022active}).
Alternative methods for FT state preparation exist~\cite{raussendorf2005quantum}, but incur a space-time trade-off and do not reduce the space-time volume.
The same considerations also apply to state preparation and logical operations in many quantum low-density parity-check (QLDPC) codes~\cite{cohen2022low,xu2024constant}.
While there have been various efforts at addressing this challenge~\cite{gottesman2013fault,yamasaki2024time}, these alternative approaches introduce higher hardware complexity~\cite{bravyi2024high,tremblay2022constant,xu2024constant,higgott2024constructions} or necessitate certain properties of the underlying codes, such as the single shot QEC property~\cite{bombin2015single,campbell2019theory,fawzi2018constant,kubica2022single,gu2024single,beverland2021cost}.

We introduce and develop an approach to FTQC that we refer to as ``transversal algorithmic fault tolerance", and show that it can lead to a substantial reduction in space-time cost for universal quantum computation.
The key idea is to consider the fault tolerance of the algorithm as a whole (Fig.~\ref{fig:fig1}(b))~\cite{delfosse2023spacetime,gidney2021stim,gottesman2022opportunities}, allowing us to produce the correct \textit{classical logical measurement output} with confidence in the presence of feed-forward operations, even though the intermediate \textit{quantum state} is far from the ideal logical state.
Contrary to the common assumption that fault-tolerant Pauli state preparation lower bounds the time cost of logical circuits by $d$~\cite{dennis2002topological,beverland2021cost,cai2023looped,duckering2020virtualized}, unless additional space-time trade-offs are made~\cite{raussendorf2005quantum}, we show that for any Calderbank-Shor-Steane (CSS) QLDPC code~\cite{steane1996error,calderbank1996good}, all transversal Clifford operations (including logical Pauli state preparation) can be performed with only constant time overhead per operation, provided that decoding can be implemented efficiently.
We verify these results through a combination of proofs and circuit-level numerical simulations of our protocol, including a simulation of state distillation factories~\cite{bravyi2005universal,fowler2012surface}, finding little change in error threshold compared to the standard surface code memory.
Specializing to the surface code, our results reduce the number of SE rounds from $\Theta(d)$ to $\Theta(1)$,
representing a direct reduction in the overall space-time overhead when classical decoding is sufficiently fast.
This is particularly relevant for neutral atoms and trapped ions, where transversal gates have been natively implemented~\cite{postler2022demonstration,ryan-anderson2022implementing,bluvstein2024logical}, and the operation timescales provide sufficient time to perform the more complex decoding problems.

\section*{Transversal Algorithmic Fault Tolerance}

\begin{figure*}
\centering
\includegraphics{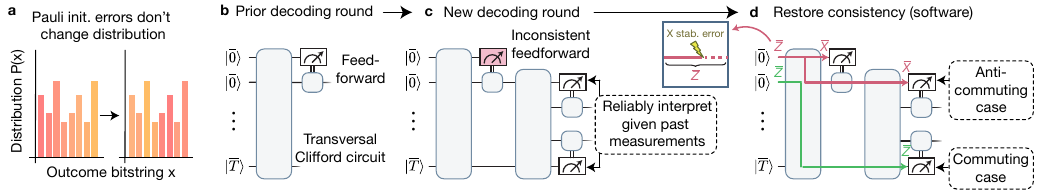}
\caption{\textbf{Illustration of fault tolerance strategy.}
(a) Illustration of the measurement distribution remaining invariant under the application of Pauli initialization errors.
(b) Logical quantum circuit with measurement and feed-forward.
All logical operations are transversal and interleaved with a single SE round, instead of $d$ SE rounds.
We must decode and commit mid-circuit to a measurement result for the top qubit, despite lacking complete syndrome information on the remaining qubits (partial decoding).
(c) With the measurement result of the top qubit, a feed-forward operation is applied and the circuit continues execution.
Upon reaching a new logical measurement, decoding is performed again on the whole circuit.
The new decoding round may assign a different result to the top qubit, causing an inconsistency in feed-forward operations.
(d) To restore consistency and reliably interpret the logical measurement results given past measurements, we apply a $\overline{Z}$ operator on the $\lket{0}$ initial state, which acts trivially on $\lket{0}$, but changes the interpreted logical measurement results to be consistent with before.
This also leads to a re-interpretation of the new logical measurement result.
Inset: Illustration of a single syndrome measurement error in the surface code, which causes a Pauli initialization error between distinct decoding rounds.
}
\label{fig:fig2}
\end{figure*}

We focus on logical circuits comprised of Clifford operations and magic state inputs, where logical Clifford operations are transversal, i.e. implemented via depth-one quantum circuits~\cite{eastin2009restrictions,jochym-oconnor2018disjointness}.
This is interleaved with SE rounds using ancilla qubits, which reveal error information on the data qubits and enable error correction.
In addition to transversal gates~\cite{shor1996fault}, we refer to preparation of data qubits in $|0\rangle$ followed by one SE round as transversal state preparation, and $Z$ basis measurement of all data qubits as transversal measurement.
To achieve universality, we allow teleporting in low-noise magic states with feed-forward operations based on past measurement results, and use the same Clifford operations above to prepare high quality magic states via magic state distillation~\cite{bravyi2005universal}.
We make use of CSS QLDPC codes, where each data or ancilla qubit interacts with a constant number of other qubits, thereby limiting error propagation, and each stabilizer generator consists of all $X$ or all $Z$ operators~\cite{calderbank1996good,steane1996error}.
Within this setting, our key result can be formulated as the following theorem:

\textbf{Theorem 1 (informal): Exponential error suppression for transversal Clifford operations with a \textit{constant} number of SE rounds per operation.}
\textit{For a transversal Clifford circuit with low-noise magic state inputs and feed-forward operations, that can be implemented with a given CSS QLDPC code family $\mathcal{Q}_d$ of growing code distance $d$, there exists a nonzero threshold $p_\mathrm{th}$, such that if the physical error rate $p<p_\mathrm{th}$ under the basic model of fault tolerance~\cite{gottesman2013fault}, then our protocol can perform logical operations with only a single SE round per operation, while suppressing the total logical error rate as $P_L=\exp(-\Theta(d))$.}

The formal theorem statement and the corresponding proof can be found in Supplementary Materials~\cite{SM}.
Following the model in Ref.~\cite{gottesman2013fault}, our proof assumes the most likely error (MLE) decoder and fast classical computation.
While this is not efficient for general hypergraph decoding problems, our simulations show that a threshold still exists for heuristic polynomial time decoders~\cite{SM}.
Fast decoders are an important area of future research, and we outline possible approaches towards this for practically-relevant circuits in the Methods section.
We consider the local stochastic noise model, where we apply depolarizing errors on each data qubit every SE round and errors on each SE measurement, with a probability that decays exponentially in the weight of the total error event.
Finally, we assume that all code patches are identical.
While this theorem assumes low-noise magic state inputs, we conjecture that for the surface code, where methods for state injection are well-developed~\cite{li2015magic,lao2022magic,gidney2023cleaner}, the same approach can be applied to certain magic state distillation procedures~\cite{bravyi2005universal} to reduce the space-time cost by a factor of $\Theta(d)$.

To understand why this result may be surprising, let us first consider the most basic task of fault-tolerantly preparing the logical $\lket{0}$ state in the surface code.
The standard procedure prepares all data qubits in $\ket{0}$, followed by $d$ SE rounds and Pauli corrections.
This allows one to gain confidence about $X$ stabilizer values, which were initially 50/50 random.

\begin{figure}
\centering
\includegraphics[width=\columnwidth]{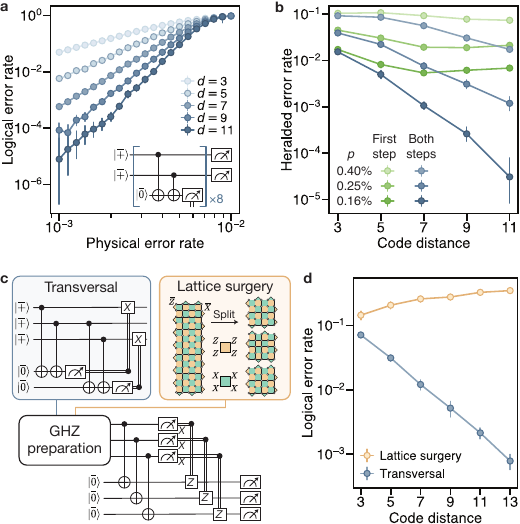}
\caption{\textbf{Numerical verification of fault tolerance.}
(a)~Simulation of circuit with repeated $\overline{ZZ}$ measurement (inset), where we commit mid-circuit to each measurement result of the logical ancilla using only the syndrome information up to that point.
The total logical error rate as a function of circuit-level physical error rate $p$, for varying code distance $d$, shows clear threshold behavior.
(b)~Probability of inconsistent assignments of logical measurement results with and without the second step of our decoding strategy, as a function of code distance and for different physical error rates, for the same circuit as (a).
Only with both steps do we observe exponential suppression of the logical error rate.
(c)~Comparison of two different methods for logical state preparation between three rotated surface codes and subsequent teleportation, for fixed circuit noise $p=0.3\%$.
We use transversal gates (left) and lattice surgery (right), in both cases with only a single SE round.
(d)~With transversal gates, the error rate decreases exponentially with the code distance.
With a single round of lattice surgery, the error rate instead increases linearly with code distance, as a single stabilizer measurement error affects the logical $\overline{ZZ}$ measurement result.}
\label{fig:fig3}
\end{figure}

If one performs only a single SE round, as our approach prescribes, then a single measurement error on the randomly-initialized $X$ stabilizers can lead to a long, erroneous chain of $Z$ corrections (inset of Fig.~\ref{fig:fig2}(d)), bringing us far from the desired logical state.
Indeed, with only $O(1)$ SE rounds and applying corrections, the resulting state is a mixture of product states and not close to any logical codeword~\cite{hastings2011topological}.
With this unreliable state preparation, we may be concerned about incorrect logical measurements and feed-forward gates~\cite{haah2024what}, which are crucial for achieving universality via magic state teleportation.
Indeed, as we numerically demonstrate in Fig.~\ref{fig:fig3}(b) and show with an explicit example in SI, new syndrome information as the circuit continues execution has a high chance of changing our assignment of a logical measurement result (Fig.~\ref{fig:fig2}(b,c)), seemingly leading to an inconsistent Clifford feed-forward operation  (Fig.~\ref{fig:fig2}(c)).
For this reason, prior work analyzing circuits with transversal gates assumed that at least $d$ SE rounds separated Pauli state initialization and measurements or out-going qubits~\cite{beverland2021cost,cai2023looped,duckering2020virtualized}.
As we discuss in Methods, Pauli state initialization is used in many important algorithmic subroutines, and the $d$ SE rounds therefore lead to a substantial cost.
Our approach overcomes this lower bound by instead considering the algorithm as a whole and exploiting the structure of initialization errors to fix these apparent inconsistencies in software.

The result of a quantum computation is fully characterized by its classical logical measurement distribution $P(z_j,z_{j-1},...,z_1)$.
Our crucial observation is that errors which do not change this distribution, such as the initialization errors described above, do not affect the output (Fig.~\ref{fig:fig2}(a)), even if they cause the intermediate quantum state to be far from an ideal logical state.

The core reason for this is that universal quantum computation is realized here by an adaptive Clifford circuit on Pauli and magic state inputs.
As the computation proceeds, Clifford feedforward gates are applied, such that the circuit executed so far is a fixed transversal Clifford circuit.
A Clifford circuit maps Paulis to Paulis, thereby mapping both Pauli initialization errors and logical Pauli stabilizers identically to other Paulis (Fig.~\ref{fig:fig2}(d)).
For each new measurement, if there is a product of measurements involving the new one that commutes with all Pauli initialization errors, then these initialization errors will not affect the product, and we can reliably and consistently infer the new measurement outcome~\cite{cain2025fast,serra-peralta2025decoding}.
If no such product exists, a Pauli error that anti-commutes with a measurement can flip its result, naively causing a logical flip.
However, the distribution is unchanged, because the anti-commutation of the measurement and corresponding logical stabilizer guarantees the result to already be 50/50 random.
At an intuitive level, the reason this always works is because an $\bar{X}$ error does not do anything on a $\lket{+}$ state.

We now provide additional elaboration of this result and show how it is concretely implemented.
We consider the joint measurement distribution of the first $j$ logical measurements (Fig.~\ref{fig:fig2}(c)):
\begin{align}
P(z_j,z_{j-1},...)=P(z_j|z_{j-1},z_{j-2},...)\cdot P(z_{j-1},z_{j-2},...),
\label{eq:conditional_distribution}
\end{align}
where $P(z_j|z_{j-1},z_{j-2},...)$ is the probability of the $j$th measurement, conditioned on previous measurement results $P(z_{j-1},z_{j-2},...)$ (Fig.~\ref{fig:fig2}(b)).
To ensure that this follows the ideal distribution, we need to guarantee two things with high confidence.
First, we must ensure that our assignment of previous measurement results $z_{j-1},\cdots,z_1$ remain unchanged, so that we maintain consistency with the feed-forward we applied (Fig.~\ref{fig:fig2}(c)).
Second, we must ensure that the conditional probability $P(z_j|z_{j-1},z_{j-2},...)$ follows the ideal conditional distribution.

By considering the algorithm as a whole and leveraging the deterministic propagation of errors through transversal Clifford circuits, one can use the surrounding syndrome history to correct for noisy measurements (Fig.~\ref{fig:fig1}(b))~\cite{cain2024correlated}.
Because the resulting decoding hypergraph (characterizing how errors triggers syndrome checks) has bounded degree, the number of connected error clusters with size $s=\Theta(d)$ is bounded~\cite{gottesman2013fault,kovalev2013fault}, limiting the probability of a logical error caused by errors that do not intersect with any state initialization.
However, we still need to account for error clusters that reach a state initialization.

More specifically, upon decoding with new syndrome information, there is a non-negligible chance that we change our assignment of the logical measurement result (green curves in Fig.~\ref{fig:fig3}(b)), seemingly leading to an inconsistent feed-forward (Fig.~\ref{fig:fig2}(c)).
Crucially, however, the logical errors that lead to these inconsistencies are structured: since magic state inputs are prepared with low-noise, they do not lead to such inconsistencies.
Therefore, the inconsistencies always correspond to a logical Pauli $X$/$Z$ operator acting on the $\lket{+}$/$\lket{0}$ initial state.
Propagating this Pauli initialization error through the executed Clifford circuit results in a Pauli product.

There are two possibilities for how this error affects a logical measurement (Fig.~\ref{fig:fig2}(d)).
First, if the measurement basis commutes with the resulting Pauli error (green), then it is unaffected and the result will be correct.
Second, if the measurement basis anti-commutes with the resulting Pauli error (red), then the result will be flipped by the error.
However, because correlated decoding ensures the correlations between current and prior measurements are correct, there is always a logical stabilizer associated with the $\lket{+}$/$\lket{0}$ initial state, which flips the logical measurements identically as the initialization error.
Applying this logical stabilizer when interpreting the logical measurement allows us to restore consistency with our previously-committed result (Fig.~\ref{fig:fig2}(d)).
This process can be efficiently determined by solving a linear system of equations~\cite{SM}.
Failure to restore consistency requires an error cluster of size $d$ that is not connected to any initialization~\cite{SM}, which has a probability exponentially small in $d$.
The above arguments show that both conditions for Eq.~(\ref{eq:conditional_distribution}) to reproduce the ideal distribution are satisfied with exponentially-vanishing failure probability, proving Thm.~1 (see SI~\cite{SM} for further details).

Specializing to the surface code and utilizing the full transversal Clifford gate set accessible to the surface code (Methods), an immediate corollary of our main theorem is a threshold result for performing constant time logical operations with an arbitrary transversal Clifford circuit, when classical decoding is sufficiently fast.
This result supports universal quantum computing when we allow magic state inputs prepared with sufficiently low noise.

Preparing high quality magic state inputs, in turn, can be performed simply with the same Clifford operations and easy-to-prepare non-fault-tolerant magic states~\cite{li2015magic,lao2022magic,gidney2023cleaner}, a procedure known as magic state distillation~\cite{bravyi2005universal}.
It will be interesting to extend our analysis to magic state distillation, where the ability to remove $d$ SE rounds from Pauli state initialization will be crucial to achieve the full savings~\cite{duckering2020virtualized,beverland2021cost,cai2023looped}.
The distillation factory and main computation can then be combined by applying our
decoding approach to the joint system.

\section*{Competitive Numerical Performance}

\begin{figure}
\centering
\includegraphics[width=\columnwidth]{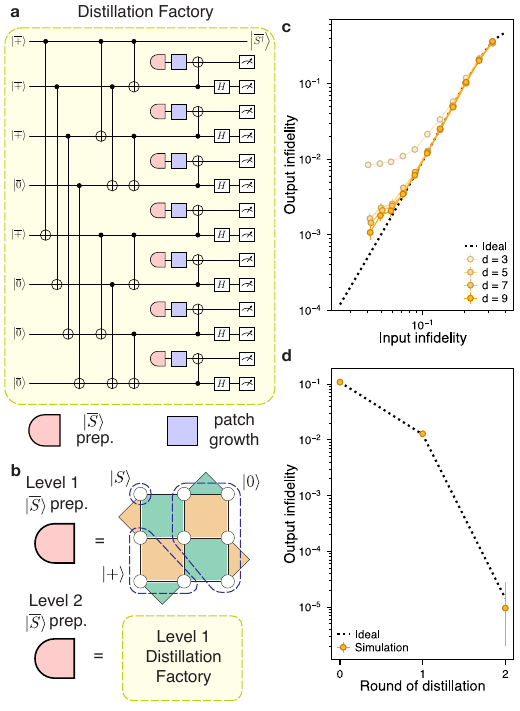}
\caption{\textbf{$\lket{S}$ state distillation factory.}
(a)~Illustration of a $\lket{S}$ state distillation factory based on the [[7,1,3]] Steane code, consisting of state initialization, layers of transversal CNOTs, followed by a teleported $\overline{S}$ gate.
Each operation involves only a single SE round.
Two of the CNOTs in the first layer act trivially and can be omitted.
(b)~The $\lket{S}$ resource state is prepared via state injection at the first level, and via the first-level factory for the second level.
(c) 1-level factory output state infidelity as a function of input state infidelity, for fixed circuit noise $p=0.1\%$ and varying levels of artificially injected $Z$ errors.
The ideal curve is calculated assuming the gate operations in the factory are perfect.
(d) Performance for one and two rounds of distillation, showing good agreement with the expected scaling.
}
\label{fig:fig4}
\end{figure}

We now turn to circuit-level simulations of our protocol to numerically evaluate its performance~\cite{gidney2021stim}, and contrast it with existing methods.
We consider various test cases of our approach that also serve as key subroutines in large-scale algorithms.

We first consider a simple circuit with intermediate logical measurements (inset of Fig.~\ref{fig:fig3}(a)) to evaluate the threshold and logical performance under partial syndrome information and without fault-tolerant state preparation.
In this example, two logical qubits are transversally initialized in $\lket{+}$, and an ancilla logical qubit is used to measure the $\overline{Z}\overline{Z}$ correlation a total of eight times, before the two logical qubits are transversally measured in the $\overline{Z}$ basis.
While individual logical measurement results are random, a correct realization of this circuit should yield the same result for $\overline{Z}\overline{Z}$ each time, despite only using syndrome information accessible up to that point.
We use the rotated surface code, a circuit-level depolarizing noise model~\cite{higgott2023improved1,cain2024correlated}, a MLE decoder based on integer programming which is not efficient in general~\cite{gurobi,cain2024correlated}, and employ the two-step process described above (see SI~\cite{SM}).

Figure~\ref{fig:fig3}(a-b) show the results of numerical simulations. 
We find that the total logical error rate shows characteristic threshold behavior, with an estimated threshold $\gtrsim 0.85\%$.
This is close to the circuit-noise memory threshold, which can be understood from the fact that an SE round involves four times as many CNOT layers as a transversal CNOT, and therefore dominates the effective error rate.
The number of SE rounds can be further optimized.
For example, in Ref.~\cite{cain2024correlated}, performing one SE round every four gate layers minimized the space-time cost per CNOT, suggesting that the practical improvement may be $\gtrsim 2d$ in some regimes~\footnote{Four transversal gates and one SE round require 8 CNOT gates in total, compared to $16d+4$ with $d$ SE rounds.}.
In Fig.~\ref{fig:fig3}(b), we further compare the scaling of heralded failure rates in the presence and absence of the second step of our decoding procedure that restores consistency, as a function of code distance $d$.
Here, the heralded failure rate is the probability that distinct rounds of decoding assign different values to the same logical measurement.
We find that this additional step is necessary to achieve exponential suppression with the code distance in our setting where the logical Pauli states are prepared with only a single SE round.

We now contrast our approach with lattice surgery in a single SE round setting~\cite{horsman2012surface,litinski2019game,kim2022fault,fowler2018low}.
We consider the logical circuit in Fig.~\ref{fig:fig3}(c), where a GHZ state preparation circuit is followed by teleportation of the GHZ state to another set of logical qubits, and then measurement in the $\overline{Z}$ basis~\cite{SM}.
Using transversal gates with only a single SE round during $\lket{+}$ and $\lket{0}$ state preparation, and decoding each logical measurement with only accessible information at that stage, we find that the logical error rate decreases exponentially with the code distance, consistent with our FT analysis.
In contrast, state preparation based on a single round of lattice surgery~\cite{kim2022fault}, which involves performing syndrome extraction with a larger code patch and then splitting it into three individual logical qubits, does not yield improved logical error rate as the code distance increases, as a single error can lead to incorrect inference of the $\overline{Z}\overline{Z}$ correlation of the GHZ state (SI).
Unlike transversal measurements, logical information here is contained in noisy stabilizer products, which require repetition to reliably infer.

Next, we simulate a state distillation factory, which involves the use of noisy state injection and is a key algorithmic subroutine for improving the quality of resource states.
We focus on distillation of the $\lket{S}=\overline{S}\lket{+}$ state using the [[7,1,3]] Steane code (Fig.~\ref{fig:fig4}(a)),
which can be efficiently simulated classically and allows the easy implementation of $\overline{S}$ gates in the surface code. 
Since this circuit has a similar structure to the practically relevant $\lket{T}$ magic state distillation factories (Extended Data Fig.~\ref{fig:t_factory}), and the state injection procedure is agnostic to the particular state that is injected, we expect the results to readily generalize to $\lket{T}$ magic state factories. 
We fix the error rate of the circuit to $p=0.1\%$, and vary the input infidelity $P_{\mathrm{in}}$ in Fig.~\ref{fig:fig4}(c).
We post-select on all $X$ logical stabilizers of the Steane code being correct to achieve cubic suppression of logical error rates (Methods).
Examining the output $\lket{S}$ of a one-level factory, we find that as the code distance is increased, the output logical error rate $P_{\mathrm{out}}$ approaches the fidelity expected for ideal Clifford logical gates in the factory $P_{\mathrm{out}}=7P_{\mathrm{in}}^3+O(P_{\mathrm{in}}^4)$ (see Methods for the full expression), across the explored fidelity regime.

Finally, we simulate the logical error rate for a two-level $\lket{S}$ state distillation factory, involving a total of 113 logical qubits, where the output $\lket{S}$ states of a $d_1=5$ factory is fed into a second factory with $d_2=9$, with the distance chosen such that the logical error is dominated by the input state infidelity $P_{\mathrm{in}}=10\%$.
As shown in Fig.~\ref{fig:fig4}(d), the logical error rates at each level of the distillation procedure are consistent with that expected based on the ideal factory formula.
We note that when performing the multi-stage distillation procedure at larger code distances, additional SE rounds may be needed to account for correlated logical errors arising from newly-initialized stabilizers in later stages (SI).

\section*{Discussion and Outlook}
Transversal operations and correlated decoding were recently found to be highly effective in experiments with reconfigurable  
neutral atom arrays~\cite{bluvstein2024logical}.
The principles of transversal algorithmic fault-tolerance described here are the core underlying mechanisms of these observations, such as correlated decoding of a logical Bell state~\cite{bluvstein2024logical}.
While recent work has provided strong evidence that this time cost reduction is possible for circuits consisting purely of Clifford gates and Pauli basis inputs~\cite{cain2024correlated}, up to now it has generally been believed that this conclusion does not hold when performing universal quantum computation~\cite{beverland2021cost,cai2023looped,duckering2020virtualized}, which crucially relies on the use of magic states and feed-forward operations.
The present work not only demonstrates that this time cost reduction is broadly applicable to universal quantum computing, but also provides a theoretical foundation for it through mathematical fault tolerance proofs.

Although our analysis focused on the use of an MLE decoder, which is inefficient in general, our numerical simulations suggest that algorithms with polynomial runtime can still achieve a competitive threshold~\cite{SM}. 
The development of improved, parallel correlated decoders and designing compilations that keep the decoding volume well-controlled is an important area of future research (see Methods for discussion).
Indeed, recent analysis~\cite{cain2025fast,serra-peralta2025decoding,turner2025scalable} shows that surface code transversal algorithms can be decoded using a fast matching decoder, without substantial increase in complexity compared to conventional decoding of individual logical qubits.
Taking into account the decoding time overhead, asymptotically, we may eventually need to insert more SE rounds (e.g. $\Theta(\log d)$ rounds for certain fast decoders~\cite{fowler2015minimum,duclos-cianci2010fast}) to simplify decoding or wait for decoding completion~\cite{terhal2015quantum}, as is also needed for FT protocols that rely on single-shot quantum error correction~\cite{bombin2015single}.
In light of recent experimental advances
~\cite{bluvstein2024logical}, a full compilation and evaluation of the space-time savings in parallel reconfigurable architectures such as neutral atom arrays is an important next step.
Finally, it will be interesting to investigate how these results can be combined with recent progress toward constant-space-overhead quantum computation~\cite{gottesman2013fault,breuckmann2021quantum,tillich2014quantum,panteleev2022asymptotically,tremblay2022constant,xu2024constant,fawzi2018constant,gu2024single} or generalized to transversal non-Clifford gates~\cite{bombin2015gauge,beverland2021cost,kubica2015unfolding,vasmer2019three,brown2020fault,bombin2018transversal,zhu2023non}, in order to further reduce the space-time volume of large-scale quantum computation.

\noindent\textbf{Acknowledgements: }
We acknowledge helpful discussions with G.~Baranes, P.~Bonilla, E.~Campbell, S.~Evered, S.~Geim, L.~Jiang, M.~Kalinowski, A.~Krishna, S.~Li, D.~Litinski, T.~Manovitz, Y.~Wu, and Q.~Xu.
We would particularly like to thank C.~Pattison for early discussions and suggesting the simulation of the $\lket{S}$ state distillation factory, and J.~Haah for stimulating discussions and deep insights.
We would also like to thank M.~Beverland, K.~Brown, U.~Vazirani and the reviewers for suggestions on presentation.
We acknowledge financial support from IARPA and the Army Research Office, under the Entangled Logical Qubits program (Cooperative Agreement Number W911NF-23-2-0219), the DARPA ONISQ program (W911NF2010021), the DARPA IMPAQT program (HR0011-23-3-0012), the DARPA MeasQuIT program (HR0011-24-9-0359), the Center for Ultracold Atoms (a NSF Physics Frontiers Center, PHY-1734011), the National Science Foundation (grant number PHY-2012023 and grant number CCF-2313084), the Army Research Office MURI (grant number W911NF-20-1-0082), DOE/LBNL (grant number DE-AC02-05CH11231), the Wellcome Leap Quantum for Bio program.
M.C. acknowledges support from Department of Energy Computational Science Graduate Fellowship under award number DE-SC0020347. 
D.B. acknowledges support from the NSF Graduate Research Fellowship Program (grant DGE1745303) and The Fannie and John Hertz Foundation.
This research was developed with funding from the Defense Advanced Research Projects Agency (DARPA). 
The views, opinions, and/or findings expressed are those of the author(s) and should not be interpreted as representing the official views or policies of the Department of Defense or the U.S. Government.

\noindent\textbf{Author Contributions: }
H.Z. formulated the decoding strategy and developed an initial proof sketch through discussions with C.Z., M.C., D.B., C.D., S.-T.W., A.K., and M.D.L..
C.Z., M.C., H.Z., and H.H. performed numerical simulations.
H.Z., A.K., C.Z., M.C., N.M., and C.D. proved the fault tolerance of the scheme.
All authors contributed to writing the manuscript.

\noindent\textbf{Data Availability: }
The data that supports the findings of this study are available at \href{https://zenodo.org/records/16552626}{https://zenodo.org/records/16552626}.

\noindent\textbf{Correspondence and requests for materials} should be addressed to H.Z. and M.D.L.\\

\bibliography{main}
\bibliographystyle{unsrt}

\clearpage

\setcounter{figure}{0}
\newcounter{EDfig}
\renewcommand{\figurename}{Extended Data Fig.}

\section{Methods}
\subsection{Background Concepts}

In this section, we review some common concepts and definitions used to establish the fault tolerance of our scheme.
We will focus on a high-level description here, and defer the formal definitions to the supplementary information.
Experienced QEC researchers may wish to skip ahead to the \textbf{key concepts} section, where we discuss a number of less commonly used concepts that are key to our results.

We start by reviewing the ideal circuits we aim to perform, based on Clifford operations and magic state teleportation.
We then describe how to turn this into an error-corrected circuit.
First, we define the local stochastic noise model that our proof assumes, which covers a wide range of realistic scenarios.
We then describe the quantum LDPC codes that we use to perform quantum error correction and how to perform transversal logical operations on them.
A noisy transversal realization of the ideal circuit is thus obtained by replacing each ideal operation by the corresponding transversal gate, followed by a single SE round.
The error-corrected realization also determines how errors trigger syndromes, which is captured in the detector error model (decoding hypergraph).
Using the detector error model and observed syndromes, we can infer a recovery operator which attempts to correct the actual errors.

Together, these concepts establish the basic procedures that are typically used for quantum error correction and conventional FT analysis.
However, in order to establish fault tolerance for our algorithmic FT protocol, we need to introduce the additional notion of frame variables, which capture the randomness of initial stabilizer projections during state preparation, and we discuss how to interpret logical measurement results in the presence of such degrees of freedom in the next section.

\textbf{Ideal circuit $\mathcal{C}$.}
We consider ideal circuits $\mathcal{C}$ in a model of quantum computation consisting of Clifford operations and magic state inputs.
$\mathcal{C}$ includes state preparation and measurement in the computational basis for any qubit, single-qubit $I$, $Z$, $H$, $S$ gates, $CNOT$ gates between any pair of qubits.
This allows the implementation of any Clifford unitary.
$\mathcal{C}$ can also include conditional operations of the above types, conditioned on previous measurement results.
Finally, $\mathcal{C}$ can also include non-Clifford magic state inputs of the form $|T\rangle=T|+\rangle$ inputs, where the $T$ gate is a $\pi/4$ rotation around the $Z$ axis.
This set of operations is known to be universal for quantum computation~\cite{nielsen2010quantum}.
We require that all qubits are measured by the end of the circuit.

\textbf{Measurement distribution $f_{\mathcal{C}}$ of ideal circuit $\mathcal{C}$.}
Ultimately, we are only interested in the classical results that our quantum computation returns.
Denote the total number of logical measurements performed throughout $\mathcal{C}$ as $M$.
The output of each execution of $\mathcal{C}$ is a bit string $\vec{b}_{\mathcal{C}}\in \mathbb{Z}_2^M$, sampled from a probability distribution $f_\mathcal{C}$.
This probability distribution fully characterizes the output of the quantum computation.

\textbf{Local stochastic noise model.} 
Our proof assumes the local stochastic noise model that is widely used in fault-tolerance analysis, see for example Ref.~\cite{gottesman2013fault}.
This noise model allows for noise correlations, but requires that the probability of any set of $s$ errors is upper-bounded by $p^s$, where $p$ is a parameter characterizing the noise strength.
We will use the local stochastic noise model in Ref.~\cite{gottesman2013fault,xu2024constant}, where the noise is applied to data qubits and the output syndrome bit.
A basis of the errors is denoted as $\mathcal{E}$ and its size scales with the space-time volume of the circuit.
For a QLDPC code (see below) and syndrome extraction circuit with bounded depth, this can be readily generalized to show a circuit-level threshold by using the fact that error propagation is bounded in a constant depth circuit~\cite{gottesman2013fault,pattison2025hierarchical,bravyi2020quantum}.

\begin{figure}
\centering
\includegraphics{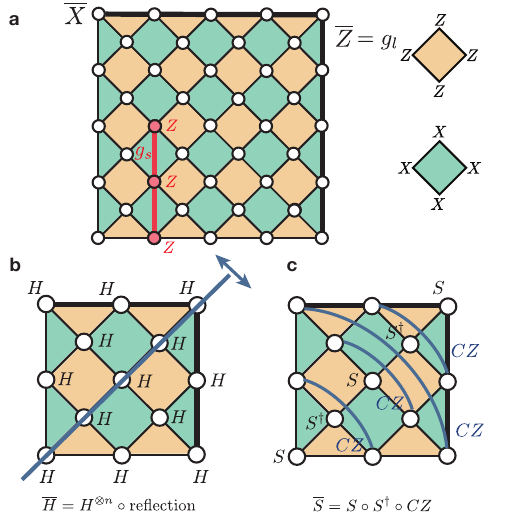}
\caption{\textbf{Surface code and transversal operations.} (a) Illustration of the surface code. White circles indicate data qubits. Orange (green) plaquettes are $Z$ ($X$) stabilizers. The logical $\overline{Z}$ ($\overline{X}$) operator runs vertically (horizontally), and we choose our convention for fixing $Z$ ($X$) stabilizers to be performing a chain of $X$ ($Z$) flips to the left (bottom) boundary, as illustrated by the red line.
(b) Illustration of transversal $\overline{H}$ gate, consisting of transversal $H$ gates followed by a reflection along the diagonal.
Note that this differs from the usual transversal $\overline{H}$ gate, which applies a rotation in the second step.
For the non-rotated surface code, both choices map $X$ ($Z$) stabilizers to $Z$ ($X$) stabilizers and hence are valid, but our choice leads to a smaller transversal partition size for the full circuit.
(c) Illustration of transversal $\overline{S}$ gate, consisting of $S$ and $S^\dagger$ gates along the diagonal, together with $CZ$ gates between mirrored qubits.}
\label{fig:surface_code_methods}
\end{figure}

\textbf{Quantum LDPC Code.}
An $[[n,k,d]]$ stabilizer quantum code $\mathcal{Q}$ is an $(r,c)$-LDPC (low-density parity check) code if each stabilizer generator has weight $\leq r$ and each data qubit is involved in at most $c$ stabilizer generators.
Here, $n$ denotes the number of phyiscal data qubits, $k$ the number of encoded logical qubits, and $d$ the code distance.
Here and below, we will use an overline to indicate logical operations and logical states, e.g. $\overline{U}$ and $\lket{0}$.
Due to the random initial stabilizer projection, we also use the separate double-bar notation $|\ooverline{0}\rangle$ to denote the ideal logical code state with all stabilizers fixed to +1.

A widely-used family of quantum LDPC codes is the surface code, due to its 2D planar layout and high threshold.
The surface code, together with its $X$ and $Z$ stabilizers and logical operators, are illustrated in Extended Data (ED) Fig.~\ref{fig:surface_code_methods}(a).

\textbf{Transversal operations.}
Consider a fixed partition of a code block, where each part contains at most $t$ qubits.
We call a physical implementation $U$ of a logical operation $\overline{U}$ transversal with respect to this partition, if it exclusively couples qubits within the same part~\cite{eastin2009restrictions,jochym-oconnor2018disjointness}.
We will also restrict our attention to the case where the logical operation, excluding SE rounds, has depth 1, motivated by the fact that the elementary gates in the ideal circuit $\mathcal{C}$ have depth 1.
We consider the same, fixed partition for all logical qubits throughout the algorithm.
This definition includes common transversal gates such as $\overline{CNOT}$ on CSS codes, for the partition where each physical qubit is an individual part.
For the surface code, we can choose a partition of size at most two, which pairs together qubits connected by a reflection.
Common Clifford operations are transversal with respect to this partition, see ED Fig.~\ref{fig:surface_code_methods}(c-d): $\overline{H}$ can be implemented via a physical $H$ on each qubit, followed by a code patch reflection in a single step.
The $\overline{S}$ gate can be implemented via $CZ$ on pairs of qubits connected by a reflection and $S$/$S^\dagger$ along the diagonal~\cite{kubica2015unfolding,moussa2016transversal,breuckmann2024fold,quintavalle2023partitioning}.
We also refer to the following state preparation and measurement in the computational basis as transversal, where $\lket{0}$ state preparation involves preparing all physical qubits in $|0\rangle$ and measuring all stabilizers once, while measurement involves measuring all physical qubits in the $Z$ basis.
Note that the $\lket{0}$ state preparation procedure does not prepare the actual code state, but rather an equivalent version with random $X$ stabilizers, where information regarding the random stabilizer initialization can be deduced later.

\textbf{Transversal realization $\tcal{C}$ of ideal circuit $\mathcal{C}$.}
If the set of operations involved in the ideal circuit (other than magic state preparation, see below) admit a transversal implementation with the QEC code $\mathcal{Q}$, then we can obtain a transversal error-corrected realization $\tcal{C}$ of the ideal circuit $\mathcal{C}$.
$\tcal{C}$ is obtained from $\mathcal{C}$ by replacing each operation by the corresponding transversal operation and inserting only one round of syndrome extraction following each gate.
Here, all transversal gate operations are Clifford gates, and non-Clifford gates are implemented via magic state teleportation.
The number of syndrome extraction rounds can be further optimized in practice~\cite{cain2024correlated}.
We denote the noiseless version of this circuit as $\tcal{C}_0$, and the circuit with a given error realization $e$ from the local stochastic noise model as $\tilde{\mathcal{C}_e}$.

The surface code provides a concrete example of a code that admits a transversal implementation of all transversal Clifford operations mentioned above.
Although we use the surface code as a concrete instance that realizes all required transversal gates, the transversal algorithmic FT construction we propose works more generally.
For a specific quantum circuit, it may be possible to compile it into, e.g. transversal CNOTs and fold-transversal gates for multiple copies of other QLDPC codes~\cite{breuckmann2024fold,quintavalle2023partitioning}, where our results also apply.

When considering magic state inputs, we assume that the magic state is initialized in the desired state with all stabilizer values fixed to $+1$, up to local stochastic noise on each physical qubit of strength $p$.
Since magic states for the surface code are typically prepared using magic state distillation, we conjecture that our methods allow single-shot logical operations during these procedures as well, which consist of Clifford operations and noisy magic state inputs (see the following section on State Distillation Factories).

\textbf{Detector error model.}
To diagnose errors, we form detectors (also known as checks), which are products of stabilizer measurement outcomes that are deterministic in the absence of errors.
A basis of detectors is denoted as $\mathcal{D}$.
We denote the set of detectors that a given error triggers as $\partial e$, which can be efficiently inferred~\cite{gidney2021stim}.
In other words, we have a linear map
\begin{align}
\partial:\mathbb{Z}_2^{|\mathcal{E}|}\rightarrow \mathbb{Z}_2^{|\mathcal{D}|}.
\end{align}
The error model, together with the pattern of detectors a given set of errors triggers, forms a decoding hypergraph $\Gamma$, also known as a detector error model, see e.g. Ref.~\cite{gidney2021stim,bombin2023modular,delfosse2023spacetime,higgott2023improved,cain2024correlated}.
The vertices of this graph are detectors, hyperedges are elementary errors, and a hyperedge is connected to the detectors that the corresponding error triggers.
During a given execution of the noisy circuit, there will be some pattern of errors $e$ that occur, giving some detection event $\partial e$.
Since the circuit is adaptive based on past measurement results, the detector error model must also be constructed adaptively to incorporate the conditional feed-forward operations.
More specifically, the decoding hypergraph $\Gamma|_j$ for the $j$th logical measurement in a given run is constructed after committing to the previous $j-1$ logical measurement results, and similarly for other objects.

To analyze error clusters, we also introduce the related notion of the syndrome adjacency graph $\Xi$~\cite{gottesman2013fault}.
In this hypergraph, vertices are elementary fault locations, and hyperedges are detectors connecting the fault locations they flip.

\textbf{Inferred recovery operator $\kappa$.}
Given the detection events and the detector error model, we can perform decoding to identify a recovery operator $\kappa\in\mathbb{Z}_2^{|\mathcal{E}|}$ which triggers the same detector pattern $\partial\kappa=\partial e.$
Our proof makes use of the most-likely-error (MLE) decoder~\cite{landahl2011fault,cain2024correlated,bravyi2015doubled}, which returns the most probable error event $\kappa$ with the same detector pattern $\partial\kappa=\partial e$.
We will refer to the combination $f=e \oplus \kappa$ as the ``fault configuration", where $\oplus$ denotes addition modulo 2.
By linearity, the fault configuration $e \oplus \kappa$ will not trigger any detectors,
\begin{align}
\partial(e\oplus \kappa)=0. 
\end{align}

\textbf{Forward-propagated error $P(e)$.}
A Pauli error $E$ occurring before a unitary $U$ is equivalent to an error $UEU^\dagger$ occurring after the unitary.
For a set of errors $e$, we can forward-propagate it through the circuit until it reaches measurements.
We denote the final operator the errors transform into as $P(e)$, and denote its restriction onto the $j$th logical measurement as $P(e)|_j$.
This is related to the cumulant defined in Ref.~\cite{delfosse2023spacetime} and the spackle operator in Ref.~\cite{bacon2014sparse}.

\begin{figure*}
\centering
\includegraphics{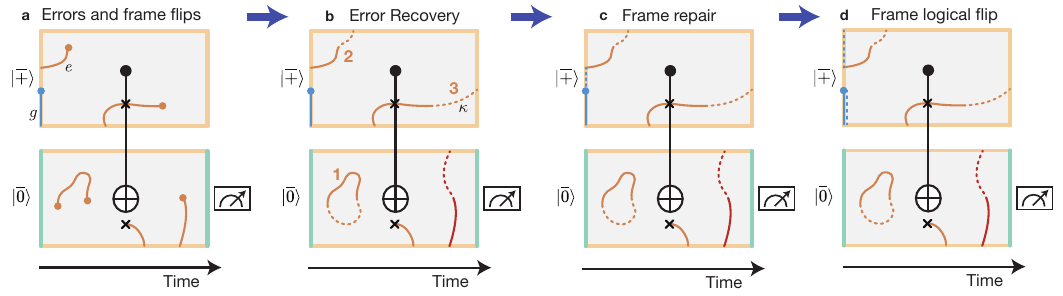}
\caption{\textbf{Illustration of error recovery and frame repair procedures.}
We illustrate the procedure for the surface code, where a cross-sectional view with one spatial axis and one time axis is shown.
We only illustrate $X$ errors and $Z$ stabilizer measurement errors, which are relevant to interpreting the $\overline{Z}$ measurement.
$X$ errors can terminate on orange boundaries, but cannot terminate on cyan boundaries.
The transversal $\overline{CNOT}$ copies $X$ errors from the top to the bottom, resulting in a branching point (black cross) and an error cluster spanning both code blocks.
(a) Error chains and frame flips. Chains of $X$-type errors (orange lines) lead to syndromes (end points) or terminate on appropriate boundaries.
A line segment in the vertical direction is a data qubit $X$ error, while a line segment in the horizontal direction is a measurement error.
Note that the $X$-type error cannot terminate on the transversal $Z$ measurement boundary. The random stabilizer initialization leads to a frame configuration on the logical $\lket{+}$ initialization, as illustrated by the blue line and the flipped $Z$ stabilizer (blue point).
This is similar to the frame stabilizer operator $g_s$ illustrated in ED Fig.~\ref{fig:surface_code_methods}(a).
(b) We first infer an error recovery operator, which has the same boundary as the error chain.
Together, the error and recovery operator form the fault configuration, which triggers no detectors.
We illustrate a few examples (orange lines) that do not lead to a logical error: (1) the fault configuration forms a closed loop and is equivalent to applying a stabilizer; (2) the fault configuration terminates on an initialization boundary; (3) the fault configuration terminates on an out-going, unmeasured logical qubit, but the forward-propagated errors onto the measured logical qubit are equivalent to a stabilizer.
A logical error can only happen when the fault configuration spans across two opposing spatial boundaries (red line), which requires an error of weight $\Theta(d)$.
(c,d) The frame repair operation returns the logical qubit to the code space with all stabilizers +1, corresponding to cancelling any residual flipped stabilizers on the initialization boundary.
Note that the error recovery process may also lead to a change that needs to be accounted for by frame repair.
An example choice of frame repair is shown in (c), which applies an overall $X$ operator on the logical measurement result.
Alternatively, a different choice of frame repair shown in (d), related to the previous one by a frame logical flip, results in identity operation on the logical measurement result.
}
\label{fig:proof_sketch}
\end{figure*}

\subsection{Key Concepts}
\label{methods:nonstandard_concepts}
We now introduce a few concepts that are less commonly discussed in the literature, but are important for our analysis.
We start by describing the randomness associated with transversal state initialization and stabilizer projections.
To do so, we introduce frame variables $g$.
To capture the random reference frame corresponding to random initialization of stabilizer values upon projection, we introduce frame stabilizer variables $g_s$.
These correspond to certain Pauli $Z$ operators that flip a subset of $X$ stabilizers, and we call both these operators and the binary vector that describes them as frame variables, where the meaning should be clear from context.
The Pauli logical initial state, e.g. $\lket{0}$, also has a logical stabilizer $\overline{Z}$, which we describe with frame logical variables $g_l$.
Applying frame logical variables on the initial state does not change the logical state, since we are applying a logical stabilizer, but this does change the interpretation of a given logical measurement shot.
To interpret logical measurement results, we must perform a frame repair operation that returns all stabilizers to +1, mirroring the error recovery inference.
However, there can be some degree of freedom in choosing the frame logical variable, which allows us to ensure consistency between multiple rounds of decoding.
These understandings lead us to propose the decoding strategy shown in Fig.~\ref{fig:fig2}, and will be crucial to our FT proofs below.

\textbf{Frame variables $g$.}
When performing transversal state initialization, all physical qubits are prepared in $|0\rangle$, and stabilizers are measured with an ancilla.
The outcome of the $X$ stabilizers will thus be random.
Following the approach taken in Ref.~\cite{gidney2021stim}, this randomness can be captured by additional $Z$ operators acting at initialization.
Concretely, for each data qubit $i$, we add $Z_i$ to a basis of frame operators $\mathcal{G}$ if it is not equivalent to any combination of operators in $\mathcal{G}$ up to stabilizers.
The state after random stabilizer projection is equivalent to starting with the ideal code state $|\ooverline{0}\rangle$ and applying a set of $Z$ operators; in other words, $|\overline{0}\rangle=g|\ooverline{0}\rangle$.
We refer to these operators as frame operators, as they describe the effective code space (``reference frame") with random stabilizers that we projected into, and help interpret logical measurement results.
The set of $Z$ operators that produces a given pattern of initial stabilizer values can be efficiently determined by solving a linear system of equations.
We choose a basis $\mathcal{G}$ for these operators, as defined above, and denote with $g$ both the Pauli operator corresponding to a frame variable as well as the binary vector describing it:
\begin{align}
g\in \mathbb{Z}_2^{|\mathcal{G}|}, \quad |\mathcal{G}|=B(n-r_Z),  
\end{align}
where $B$ is the number of code blocks used, $n$ is the number of data qubits per block and $r_Z$ is the number of independent $Z$ stabilizer generators per block.
In the presence of noise, we can imagine first performing the random stabilizer projection perfectly, and then performing a noisy measurement of the syndromes via ancillae and recording the results.
Although this does not allow the reliable inference of frame variables, we will show that the transversal measurement provides enough information to infer the relevant degrees of freedom for interpreting logical measurement results.

\textbf{Frame logical variables $g_l$.}
A special subset of frame variables are frame logical variables
\begin{align}
g_l\in\mathbb{Z}_2^{Bk},
\end{align}
which are combinations of the $Z$ operators that form a logical $\overline{Z}$ operator of the code block, and therefore act trivially on the code state $\lket{0}$.
Here, $B$ is the number of code blocks and $k$ is the number of logical qubits per block.
While they do not change the initialized physical state, nor do they flip any stabilizers, different choices of the frame logical variables when decoding will lead to different interpretations of the logical measurement result, as we explain next.

\textbf{Frame stabilizer variables $g_s$.}
We refer to frame variables that are not frame logical variables as frame stabilizer variables.
These variables will flip the randomly initialized stabilizer values.
An example is shown in ED Fig.~\ref{fig:surface_code_methods}(a), in which a chain of $Z$ errors connecting to the bottom boundary flips a single stabilizer.

\textbf{Interpreting logical measurement outcomes in the presence of frame variables.}
We now describe how to interpret logical measurement results in the presence of randomly initialized frame variables.

First, in the presence of noise, we apply the decoding procedure and obtain an error recovery operator $\kappa$ such that $\partial(\kappa\oplus e)=0$.
Note that $\kappa\oplus e$ may have some nontrivial projection onto the initialization boundary, such as string 2 that terminates on the $\lket{+}$ boundary in ED Fig.~\ref{fig:proof_sketch}(b).
This projection can modify the effective frame, and must be taken into account when returning things to the code space.

Next, we perform an analogous procedure to error recovery for the frame variables.
Specifically, we perform a frame repair operation
\begin{align}
\lambda\in \mathbb{Z}_2^{|\mathcal{G}|}
\end{align}
to return to the code space with all stabilizers set to +1.
This corresponds to an inference of what the reference frame was after the random stabilizer projection during initialization, and the repair operation should be viewed as being applied on the corresponding initialization boundary as well.
In other words, we require $(e\oplus\kappa)\oplus(g\oplus\lambda)$ to act as a stabilizer or logical operator, such that the stabilizer values are the same as the ideal code state $|\ooverline{0}\rangle$.
We will refer to the combination $h=g\oplus\lambda$ as the ``frame configuration".
Following this step, all frame stabilizer variables $g_s$ have been determined, but we still have freedom to choose our frame logical variables $g_l$.

Finally, we evaluate the product of Pauli operators to determine the logical measurement result.
Denote the raw logical observable inferred from the bit strings as
\begin{align}
\overline{L}(z)=\bigoplus_{z_i\in \overline{L}} z_i,
\end{align}
and the corrected logical observable after applying the error recovery operation $\kappa$ and frame repair operation $\lambda$ as \begin{align}
\overline{L_c}(z,\kappa,\lambda)=\overline{L}(z)\oplus \overline{F}(\kappa)\oplus \overline{F}(\lambda),
\end{align}
where $\overline{F}(\kappa)$, $\overline{F}(\lambda)$ indicates the parity flip of the logical observable due to the operator $\kappa$, $\lambda$.

In the noiseless case, the raw logical measurement result is equivalent to the ideal measurement result that one would obtain if one had perfectly prepared the ideal code state $|\ooverline{0}\rangle$, up to the application of $\overline{F}(g\oplus \lambda)$ on the initial state.
However, $g\oplus \lambda$ consists of physical $Z$ operations only and commutes with all stabilizers, so it must act as a combination of $Z$ stabilizers and logical $Z$ operator on $|\ooverline{0}\rangle$.
Therefore, it does not change the distribution of measurement results, although it can change the interpretation of individual shots.
The procedure in the noisy case can be reduced to the noiseless case after applying the MLE recovery operator $\kappa$, with a suitable modification to the repair operation $\lambda$ to account for fault configurations that terminate on initialization boundaries and therefore forward-propagated to flip some stabilizers on the relevant logical measurement (ED Fig.~\ref{fig:proof_sketch}(c)).

\textbf{Decoding strategy.}
A key component of our FT construction is the decoding strategy.
In our setting with transversal Clifford gates only, classical decoding only becomes necessary when we need to interpret logical measurement results.
We sort the set of logical measurements into an ordering $\{\bar{L}_1, \bar{L}_2, \bar{L}_3,...,\bar{L}_M\}$ based on the time they occur, and then decode and commit to their results in this order.

For the $j$th logical measurement $\bar{L}_j$, we first apply the most-likely-error (MLE) decoder to the available detector data $\mathcal{D}|_j$ and the detector error model $\Gamma|_j$, where $|_j$ denotes that this information is restricted to information up to the $j$th logical measurement.
Note that since we allow feed-forward operations, the decoding hypergraph may differ in each repetition of the circuit (shot).
After this first step, we will have obtained an inferred recovery operator $\kappa$, similar to standard decoding approaches.

The second step is to apply frame logical variables $g_l$ such that previously-committed logical measurement results retain the same measurement result.
It may not be clear a priori that this is always possible, but we prove that below a certain error threshold $p_{th}$, the probability of a failure decays to zero exponentially in the code distance.
This guarantees that we are always consistently assigning the same results to the same measurement in each round of decoding.
The assignment of frame logical variables can be solved efficiently using a linear system of equations.

\subsection{Proof Sketch}
\label{methods:proof_sketch}
In this section, we provide a sketch of our FT proof, using the concepts introduced above.
The detailed proofs are provided in the SI, and we reference the section they correspond to in the SI for interested readers.
Our reasoning follows three main steps:
\begin{enumerate}
\item We show that the transversal realization reproduces the logical measurement result distribution of the ideal circuit, regardless of the reference frame we initially projected into.
\item We obtain perfect syndrome information on the logical qubits via transversal measurements, which we then combine with correlated decoding to handle errors throughout the circuit and guarantee that any logical error must be caused by a high-weight physical error cluster.
\item By counting the number of such high-weight error clusters, we show that when the physical error probability is sufficiently low, the growth in the number of error clusters as the distance increases is slower than the decay of probability of high-weight clusters, thereby establishing an error threshold and exponential sub-threshold error suppression.
\end{enumerate}

We now explain a set of useful lemmas that lead to our main theorem.

\textbf{Frame variables $g$ do not affect the logical measurement distribution (Lemma 4 of SI).}
We show that the choice of frame variables $g$ does not affect the logical measurement distribution $f_{\tilde{C}}$.
Intuitively, this is because different choices of frame variables are equivalent up to the application of $\bar{Z}$ logicals on $|\ooverline{0}\rangle$, which does not affect the logical measurement distribution.
Indeed, as long as we are able to keep track of which subspace of random stabilizer values we are in, achieved via the transversal measurement, the measurement result distribution should not be affected.

\textbf{$f_{\mathcal{C}}=f_{\tcal{C}_0}$ (Sec. II.6 of SI).}
In other words, the noiseless transversal realization $\tcal{C}_0$ produces the same distribution of logical bit strings as the ideal quantum circuit $\mathcal{C}$.
This can be seen from the previous statement by choosing all frame variables to be zero and invoking standard definitions of logical qubits and operations.

\textbf{Transversal gates limit error propagation (Lemma 6 of SI).}
One major advantage of transversal gates is that they limit error propagation~\cite{shor1996fault,gottesman2010introduction}, thereby limiting the effect any given physical error event can have on any logical qubit.
With the bounded cumulative partition size $t$ defined above, one can readily show that any error $e$ acting on at most $k$ qubits can cause at most $tk$ errors on a given logical qubit, when propagated to a logical measurement $P(e)|_j$.

\textbf{Effect of low-weight faults on code space (Lemma 7 of SI).}
Consider the syndrome adjacency graph $\Xi|_j$, which is the line graph of the detector error model $\Gamma|_j$ corresponding to the first $j$ logical measurements, and any fault configuration $f|_j=(e\oplus\kappa)|_j$.
We show that if the largest weight of any connected cluster of $f|_j$ is less than $d/t$, then there exists a choice of frame repair operator $\hat{\lambda}_j$, such that the forward propagation of fault configuration and frame configuration
\begin{align}
P(e|_j\oplus\kappa|_j)\oplus P(g|_j\oplus \hat{\lambda}_j)
\end{align}
acts trivially on the first $j$ logical measurements.

The intuition for this statement is illustrated in ED Fig.~\ref{fig:proof_sketch}.
Suppose without loss of generality that the logical measurement we are examining is in the $Z$ basis, then we only need to examine errors that forward-propagate to $X$ errors.
By definition, the fault configuration $e\oplus \kappa$ and frame configuration $g\oplus \lambda$ should return things to the code space and not trigger any detectors, implying that the $X$ basis component of \mbox{$P(e\oplus \kappa\oplus g\oplus\lambda)=P(f\oplus h)$} is a product of $X$ stabilizers and logical operators.
Consider each connected component $f_i$ of $f|_j$, then by transversality (previous lemma) and $\textrm{wt}(f_i)<d/t$, we have $\textrm{wt}(P(f_i))<d$.

Case 1: If $f_i$ does not connect to a Pauli initialization boundary (fault configurations 1 and 3 in ED Fig.~\ref{fig:proof_sketch}(b)), then it is also a connected component of $f\oplus h$, since the frame configuration lives on the initialization boundary.
Since $P(f_i)$ has weight less than $d$, it must be a stabilizer and therefore acts trivially on the logical measurement under consideration.

Note that because magic states are provided with known stabilizer values up to local stochastic noise, connected components of the fault configuration cannot terminate on them without triggering detectors.
The same also holds for measurement boundaries or boundaries in which the initialization stabilizer propagates to commute with the final measurement.
Only when the initialization stabilizer propagates to anti-commute can we connect to the boundary, as described in case 2, but this also then implies that the measurement is 50/50 random and can be made consistent using our methods.

Case 2: Now suppose $f_i$ connects to an initialization boundary (fault configuration 2 in ED Fig.~\ref{fig:proof_sketch}(b)) and its connected component $P(f_i\oplus h_i)$ acts as a nontrivial logical operator $L$, flipping the logical measurement.
In this case, we can choose a different frame repair operator such that $P(\hat{\lambda})=P(\lambda)\oplus L$, which does not flip the logical measurement.
In ED Fig.~\ref{fig:proof_sketch}(c,d), we can intuitively think of this as changing whether the frame repair connects in the middle or to the two boundaries.
In one of these two cases, the total effect of the fault configuration and frame configuration is trivial on the logical measurements of interest (ED Fig.~\ref{fig:proof_sketch}(d) in this case).

Thus, we see that when the fault configuration only involves connected clusters of limited size, its effect on the logical measurement results is very limited.
This leads to a key technical lemma that lower bounds the number of faults required for a logical error to occur.

\textbf{Logical errors must be composed of at least $d/t$ faults (Lemma 8 of SI).}
Due to the decoding strategy we employ, there are two types of logical errors we must account for.

First, we may have a logical error in the usual sense, where the distribution of measurement results differs from the ideal quantum circuit $f_{\tcal{C}}\neq f_{\mathcal{C}}$.
It is important to note here, however, that this deviation is in the distribution sense.
Thus, if a measurement outcome that was 50/50 random was flipped, it does not cause a logical error yet, as the outcome is still random.
In this case, it is only when the joint distribution with other logical measurements is modified that we say a logical error has occurred.
When analyzing a new measurement result with some previously committed results, we analyze the distribution conditional on these previously committed results.

Second, there may be a heralded logical error, in which no valid choice of frame repair operation $\lambda$ exists in the second step of our decoding strategy.
More specifically, there is no $\lambda$ that makes all logical measurement results identical to their previously-committed values.

We show that when the largest weight of any connected cluster in the fault configuration is less than $d/t$, neither type of logical error can occur.
The absence of unheralded logical errors can be readily seen from the above characterization of the effect of low-weight faults on the code space.
To study heralded errors, we make slight modifications to analyze the consistency of multiple rounds of decoding, and find that heralded errors require one of the two rounds of decoding that cannot be consistently assigned to have a fault configuration with weight $\geq d/t$, thereby leading to the desired result.

\textbf{Counting lemma (Lemma 5 of Ref.~\cite{gottesman2013fault}).}
The counting lemma is a useful fact that bounds the number of connected clusters of a given size within a graph, with many previous uses in the QEC context~\cite{aliferis2007accuracy,bombin2015single,kovalev2013fault,gottesman2013fault,kubica2022single}.
It shows that for a graph with bounded vertex degree $v$ and $n$ vertices, as is the case for the syndrome adjacency graph $\Xi$ of qLDPC codes, the total number of clusters of size $s$ is at most $n(ve)^{s-1}$.
This bounds the number of large connected clusters.
When the error rate is low enough, the growth of the ``entropy" factor associated with the number of clusters will be slower than the growth of the ``energy" penalty associated with the probability, and thus the logical error rate will exponentially decrease as the system size is increased, allowing us to prove the existence of a threshold and exponential sub-threshold suppression.

\textbf{Theorem 1: Threshold theorem for transversal realization $\tcal{C}$ with any CSS QLDPC code, with reliable magic state inputs and feed-forward (Theorem 10 of SI).}
With the preceding lemmas, we can prove the existence of a threshold under the local stochastic noise model.
Using the counting lemma, we can constrain the number of connected clusters $N_s$ of a given size $s$ on the syndrome adjacency graph $\Xi$.
For a connected cluster of size $s$, MLE decoding implies that at least $s/2$ errors must have occurred, which has bounded probability scaling as $p^{s/2}$ under the local stochastic noise model.
Our characterization of logical errors implies that a logical error can only occur when $s\geq d/t$.
For each round of logical measurements, the probability of a logical error is then bounded by a geometric series summation over cluster sizes $s$, with an entropy factor from cluster number counting and an energy factor from the exponentially decreasing probability of each error event:
\begin{align}
P_{err}&\propto M\sum_{s=\frac{d}{t}}^{\infty}N_s 2^s p^{s/2}\nonumber\\
&\propto \qty(2v\e\sqrt{p})^{d/t}
=\qty(\frac{p}{1/(2v\e)^2})^{d/2t},
\end{align}
where $v$ is a bound on the vertex degree of the syndrome adjacency graph and is dependent on the degrees $r$ and $c$ of the QLDPC code.
When the error probability $p$ in the local stochastic noise model is sufficiently small, the latter factor outweighs the former, and the logical error rate decays exponential to zero as the code distance increases, with an exponent $p^{d/2t}$.
We can then take the union bound over rounds of logical measurements to bound the total logical error probability.

While our theorem assumes reliable magic state inputs with local stochastic data qubit noise only, we expect our results to readily generalize to magic state distillation factories (see next section and discussion in main text), thereby enabling a $\Theta(d)$ saving for universal quantum computing.

Note that to prove a threshold theorem for FT simulating the ideal circuit $\mathcal{C}$, we need a family of codes $\{\mathcal{Q}\}$ with growing size that provide a transversal realization of $\mathcal{C}$.
For general high-rate QLDPC codes, this may be challenging, as the set of transversal gates is highly constrained~\cite{breuckmann2024fold,quintavalle2023partitioning}.
However, we will now show that the surface code provides the required code family.

\textbf{Theorem 2: Fault tolerance for arbitrary Clifford circuits with reliable magic state inputs and feed-forward, using a transversal realization with the surface code (Theorem 11 of SI).}
We can further specialize the preceding results to the case of the surface code.
With the transversal gate implementations of $H$, $S$ and $CNOT$, we can implement arbitrary Clifford operations with cumulative partition size $t=2$.
Note that with more detailed analysis of the error events and gate design, it may be possible to recover the full code distance $d$ (instead of the $d/2$ proven here), which we leave for future work.
Our threshold and error suppression results are independent of the circuits implemented, e.g. whether the circuit has a large depth or width.
The resulting logical error rate scales linearly with the circuit space-time volume and number of logical measurements, and is exponentially suppressed in the code distance, similar to the usual threshold theorems.

A straight-forward application of the previous theorem shows the existence of a threshold and exponential sub-threshold error suppression.
Importantly, the surface code provides all elementary Clifford operations, thereby giving a concrete code family for the FT simulation of any ideal circuit $\mathcal{C}$, as long as we are provided with the appropriate magic state inputs, which can in turn be obtained in the same way via magic state distillation.

\subsection{State Distillation Factories}
In this section, we provide more details on state distillation factories.
First, we derive the output fidelity of the $\lket{Y}$ state distillation factory described in the main text, as a function of input $\lket{Y}$ state fidelity and assuming ideal Clifford operations within the factory.
Second, we illustrate the 15-to-1 $\lket{T}$ magic state distillation factory and comment on a few simplifications that our decoding strategy enables in executing this factory.

\begin{figure}
\centering
\includegraphics{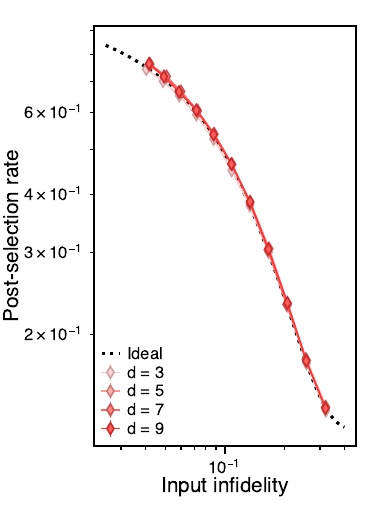}
\caption{\textbf{Post-selection rates from the numerical simulation of $\lket{S}$ state distillation factory at circuit noise $p=0.1\%$.} Here the ideal post-selection rates are defined in Eq.~\eqref{eq:s-post-selection-rate}.}
\label{fig:s-post-selection-rate}
\end{figure}

The $\lket{Y}$ state distillation factory described in the main text prepares a Bell pair between a single logical qubit and seven logical qubits further encoded into the $[[7,1,3]]$ Steane code.
Applying a transversal $\overline{S}$ gate on the Steane code then leads to a $\overline{S}$ gate on the output logical qubit due to the Bell pair.
Error detection on the Steane code further allows one to distill a higher-fidelity logical state.
For this distillation factory, we can directly count the error cases for the magic state input that lead to a logical error, conditional on post-selection results.
For example, there are seven logical $\overline{Z}$ representatives of weight three and one logical representative of weight seven, and the application of a logical representative leads to an undetectable error.
Counting all possible combinations, we arrive at the following formula for noisy magic state inputs and ideal Clifford operations
\begin{align}
P_{\mathrm{out}}&=\frac{7P_{\mathrm{in}}^3(1-P_{\mathrm{in}})^4+P_{\mathrm{in}}^7}{(1-P_{\mathrm{in}})^7+7P_{\mathrm{in}}^3(1-P_{\mathrm{in}})^4+7P_{\mathrm{in}}^4(1-P_{\mathrm{in}})^3+P_{\mathrm{in}}^7}\nonumber\\
&\approx 7P_{\mathrm{in}}^3,
\end{align}
where $P_{\mathrm{out}}$ is the output logical error rate and $P_{\mathrm{in}}$ is the input logical error rate. For our numerical simulations, we artificially inject $Z$ errors for the input state. The denominator represents the post-selection rate from the distillation factory
\begin{equation}
    P_{ps} = (1-P_{\mathrm{in}})^7+7P_{\mathrm{in}}^3(1-P_{\mathrm{in}})^4+7P_{\mathrm{in}}^4(1-P_{\mathrm{in}})^3+P_{\mathrm{in}}^7.
    \label{eq:s-post-selection-rate}
\end{equation}
From the numerical simulation of the $\lket{S}$ state distillation factory (Fig.~\ref{fig:fig4}) using our decoding strategy, we also confirm that the post-selection rates match theoretical ones. Here, post-selection happens at different stages. 
\begin{itemize}
    \item At the injection stage, we first grow a physical qubit into a surface code patch with distance $d_0=3$ by measuring all stabilizers for two rounds. If no error is detected from these two rounds of syndrome extraction, we say that this shot passes the \textit{injection test}. After passing the injection test, the patch will be grown into distance $d$ by measuring stabilizers at new distance for one round without post-selection.
    \item At the distillation stage, we only have syndrome information prior to measuring out the output distilled $\lket{S}$ state, and decode logical $\overline{X}$ measurements to determine whether to keep or discard this shot. If $X$ stabilizers of the $[[7,1,3]]$ code reconstructed from the decoded logical $\overline{X}$ measurements are $+1$, we say that this shot passes the \textit{factory test}.
\end{itemize}
In ED Fig.~\ref{fig:s-post-selection-rate}, we plot the factory test post-selection rate as a function of the input error rate $P_{\mathrm{in}}$, representing the post-selection rate coming from the operation of the magic state factory. The input infidelity is estimated by simulating the state injection protocol itself at $d_0=3$ and different $d$ with the same $p$ and $P_{\mathrm{in}}$.

\begin{figure}
\centering
\includegraphics{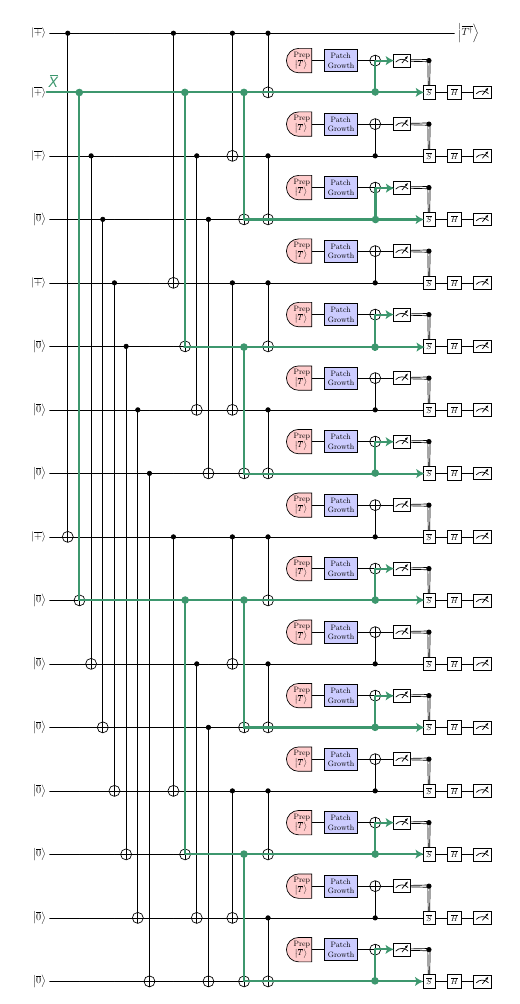}
\caption{\textbf{Illustration of a 15-to-1 $\lket{T}$ magic state distillation factory}, adapted from Ref.~\cite{beverland2021cost}. The green lines illustrate the application of a logical stabilizer, which allows re-interpretation of measurement results and changes which feed-forwards should be performed.}
\label{fig:t_factory}
\end{figure}

In ED Fig.~\ref{fig:t_factory}, we illustrate the 15-to-1 $\lket{T}$ state distillation factory, which takes 15 noisy $\lket{T}$ states and distills a single high quality $\lket{T}$ state.
As described in Ref.~\cite{bravyi2005universal}, assuming ideal Clifford operations, the rejection probability scales linearly with the input infidelity, while the output logical error rate scales with the cube of the input infidelity.
The $\lket{T}$ factory bears a lot of similarities with the $\lket{S}$ factory in the main text: In both cases, we start with Pauli basis states, apply parallel layers of CNOT gates, and then perform resource state teleportation using a CNOT.
The resource states at the lowest level can be prepared using state injection, which is agnostic to the precise quantum state being injected and therefore should apply equally to a $\lket{S}$ and $\lket{T}$ state, while the resource states at the higher levels are obtained by lower levels of the same distillation factory.
The main difference is that because the feed-forward operation is now a Clifford instead of a Pauli, the feed-forward gate must be executed in hardware, rather than just kept track of in software.

When performing magic state distillation and teleporting the magic state into the main computation, the first step of our protocol requires correlated decoding of the distillation factory and main computation together.
It will be interesting to formally extend our threshold analysis to incorporate noisy magic state injection and state distillation procedures.

Using our decoding strategy, it is possible to reduce the number of feed-forward operations that need to be executed.
As illustrated in ED Fig.~\ref{fig:t_factory}, we can apply an $\overline{X}$ operator on the $\lket{+}$ logical initial states, which is a logical stabilizer of the resulting quantum state.
Applying this operator flips the interpreted results of some subset of logical measurements.
Thus, we can always choose to not apply a feed-forward $\overline{S}$ on the first $\lket{T}$ teleportation, but instead change what feed-forward operations are applied on the remaining $\lket{T}$ teleportations.
There are 15 $\lket{T}$ teleportations to be implemented and 5 $\lket{+}$ logical state initialization locations.
Therefore, we expect that at most 10 feed-forward operations need to be applied.
Using these techniques, the logical qubit locations where the feed-forward operations need to be applied may also be adjusted, which may be beneficial for the purpose of control parallelism~\cite{bluvstein2024logical}.

Finally, we also comment on the relation of our results to other computational models that make use of magic state inputs and Clifford operations.
In particular, Pauli-based computation~\cite{bravyi2016trading,yoganathan2018quantum} has been shown to provide a weak simulation of universal quantum circuits using only magic state inputs, apparently removing the need of $\lket{0}$ and $\lket{+}$ logical states altogether, and clarifying the importance of $\lket{T}$ state preparation in particular.
However, this model relies on the logical measurements being non-destructive, and continues to use a given logical qubit after measurement, which is not possible for transversal measurements on logical qubits without Pauli basis initialization.
Thus, in an error-corrected implementation, Pauli basis initialization is still necessary, and the use of our FT framework is necessary to achieve low time overhead.
This comparison to other computational models highlights the generality of the algorithmic fault-tolerance framework, and indicates that universally across these various computational models, such techniques allow a $\Theta(d)$ saving.

\subsection{Importance of Shallow Depth Algorithmic Gadgets}
\label{methods:shallow_gadget}
In this section, we discuss the importance of shallow-depth algorithmic gadgets in many practical compilations of quantum algorithms.
This highlights the need for FT strategies that do not require a $\Theta(d)$ separation between initialization and measurement, as we developed in the main text.

In general, circuit components that involve an ancilla logical qubit often have a shallow depth between initialization and measurement, whether this ancilla is used for algorithmic reasons or compilation reasons.
For instance, temporary ancilla registers are used in algorithmic gadgets such as adders~\cite{gidney2018halving,cuccaro2004new} or quantum read-only memories~\cite{babbush2018encoding}, where the bottom rail of a ripple carry structure is initialized, two or three operations are performed on it, and then the ancilla qubit is measured.
A useful technique for performing multiple circuit operations in parallel is time-optimal quantum computation~\cite{fowler2012time,gidney2019how,litinski2022active}, which is also related to gate teleportation~\cite{nielsen2010quantum} and Knill error correction~\cite{knill2004quantum}.
In this case, a pair of logical qubits are initialized in a Bell state.
One qubit is then sent as the input into a circuit fragment $A$, while the other qubit executes a Bell basis measurement with the output of another circuit fragment $B$.
The combined circuit is equivalent to the sequential execution of $B$ and $A$.
This allows the two circuit fragments to be executed in parallel, despite them originally being sequential, thereby reducing the total circuit depth and idling volume.
However, to fully capitalize on this advantage, it is desirable to only have a constant number of SE rounds separating the Bell state initialization and Bell basis measurement, in order to minimize the extra circuit volume incurred by the space-time trade-off.
Thus, a depth $O(1)$ separation between state initialization and measurement is again highly desirable.

Another common situation in which there is a low-depth separation between initialization and measurement is magic state distillation~\cite{bravyi2005universal} and auto-corrected magic state teleportation~\cite{gidney2019flexible}.
Many magic state factories involve a constant-depth Clifford circuit  (e.g. depth 4 for the 15-to-1 distillation factory), followed by application of non-Clifford rotations~\cite{bravyi2005universal,beverland2021cost,fowler2012surface,bravyi2012magic}.
The non-Clifford rotations are often implemented via noisy magic states and gate teleportation, which therefore require logical measurements.
If the Clifford circuit depth has to be at least $d$ to maintain FT, as is assumed in e.g. Ref.~\cite{cai2023looped}, the time cost of the magic state factory will be much larger than the case in which we can execute the circuit fault-tolerantly in constant depth, as we demonstrate here.

\subsection{Decoding Complexity}
\label{methods:decoding}

In this section, we discuss the decoding complexity of our FT construction, and highlight important directions of future research.
While a detailed analysis and high-performance implementation of large-scale decoding is beyond the scope of this work, this will be important for the large-scale practical realization of our scheme and to maximize the savings in space-time cost.
We therefore sketch some key considerations and highlight important avenues of research that can address the decoding problem.
We emphasize that much of our discussion is not specific to our FT strategy, and may also apply to other hypergraph decoding problems and existing discussions of single-shot QEC~\cite{bombin2015single} (Supplementary Information).

Compared with usual decoding problems, there are two main aspects that may increase the complexity in our setting.
First, the decoding problem is now by necessity a hypergraph decoding problem, involving hyperedges connecting more than two vertices, which are not decomposable into existing weight-two edges~\cite{cain2024correlated}.
Second, the size of the relevant decoding problem (decoding volume) may be much larger, as one needs to jointly decode many logical qubits, in the worst-case increasing exponentially with the code distance up to the scale of the full system.

The hypergraph decoding problem has been studied in a variety of different settings~\cite{cain2024correlated,acharya2023suppressing,wootton2012high,fowler2013optimal,delfosse2022toward}, and heuristic decoders appear to handle this fairly well in the low error rate regime in practice.
For example, polynomial-time decoders such as belief propagation + ordered statistic decoding (BPOSD)~\cite{panteleev2019degenerate}, hypergraph union find (HUF)~\cite{delfosse2022toward,cain2024correlated}, and minimum-weight parity factor (MWPF)~\cite{wu2024hypergraph} have been shown to result in competitive thresholds.
Decoding on hypergraphs is also often required for high-rate QLDPC codes, or to appropriately handle error correlations.
Fully addressing the hypergraph decoding problem in the context of logical algorithms is an important area of future research.

There are several ways in which the increased decoding volume can be dealt with.
First, for many key subroutines, the algorithmic structure may lead to a much slower growth of the relevant circuit volume for decoding.
For example, the main subroutines involved in state-of-the-art factoring algorithms~\cite{gidney2019how}---quantum addition via a ripple-carry adder and quantum look-up tables---both have a structure in which the number of relevant qubits grows only linearly with the depth, rather than exponentially.
Combined with the fact that error inferences that are sufficiently far $\Omega(d)$ away from measurements or out-going qubits can be committed to without affecting the logical error rate~\cite{bombin2023modular}, the total relevant decoding volume remains polynomial in the code distance in many important algorithm implementations.
Moreover, for underlying codes with the single-shot QEC property~\cite{bombin2015single}, it may be possible to further reduce the relevant volume.

Second, extra QEC rounds can also be inserted to reduce the relevant decoding volume and give more time for the classical decoder to keep up with the quantum computer and avoid the backlog problem~\cite{terhal2015quantum}.
Asymptotically, this may be necessary for both our scheme and for computation schemes based on single-shot quantum error correction~\cite{bombin2015single,bombin2013gauge}, unless $O(1)$-time classical decoding is possible.
In both cases, the time cost will grow from $\Theta(1)$ to $\Theta(d/C)$, where the  improvement factor $C$ over conventional schemes with $d$ SE rounds can be made arbitrarily large as the classical computation is sped up.

Third, we expect algorithms based on cluster growth (HUF and MWPF) and belief propagation to be readily amenable to parallelization across multiple cores~\cite{liyanage2023scalable,richardson2008modern,fowler2015minimum,wu2023fusion}, with the decoding problems merging only when an error cluster spans multiple decoding cores.
As an error cluster of size $\Theta(d)$ is exponentially unlikely, we expect it to be unlikely for many decoding problems to have to be merged together.
Indeed, fast parallel decoders for the surface code~\cite{fowler2015minimum,wu2023fusion} and QLDPC codes~\cite{grospellier2019} have been argued to achieve average runtime $O(1)$ per SE round, although they still have an $O(d)$ or $O(\log d)$ latency.
Therefore, although the original decoding problem is not modular (input-level modularity)~\cite{tan2023scalable,skoric2023parallel,bombin2023modular}, in practice we may expect the decoder to naturally split things into modular error clusters (decoder-level modularity).

Finally, there are many additional optimizations that can be applied in practice.
Because the decoding problems have substantial overlap, it may be possible to make partial use of past decoding results, particularly when using clustering decoders.
The decoding and cluster growth process can also be initiated with partial syndrome information and continuously updated as more information becomes available.
Decoding problems with specific structure, such as circuit fragments in which the flow of CNOTs are directional (ED Fig.~\ref{fig:t_factory}), may also benefit from specialized decoders~\cite{beverland2021cost}.
We also note that although the relevant decoding hypergraph for any given measurement is now larger, for a given rate of syndrome extraction on the hardware, the amount of incoming data is comparable to the usual FT setting.
Although the individual correlated decoding problem is larger, we will only need to solve very few of them.
In both algorithmic FT and conventional FT, we expect the total amount of classical decoding resources to scale with the number of logical qubits.
When decomposing correlated decoding into individual cluster decoding problems, we therefore expect the aggregate classical decoding resources required for our protocol to still remain competitive with conventional approaches.

\subsection{Hardware Considerations}
\label{methods:hardware}
In this section, we briefly comment on the hardware requirements to implement our scheme.
It is worth emphasizing that these requirements may be relaxed with future improvements to our construction.

Our algorithmic FT protocol makes important use of transversal gate operations between multiple logical qubits.
As such, a direct implementation likely requires two key ingredients: long-range connectivity and reconfigurability.
Long-range connectivity is used to entangle physical qubits that are located at matching positions in large code patches, which are otherwise spatially-separated.
Reconfigurability is useful because a given logical qubit may perform transversal gates with many other logical qubits throughout its lifetime, such that a high cumulative connectivity degree is required, or multiple swaps and routing must be used.
Given that common routing techniques based on lattice surgery incur a $\Theta(d)$ time cost, it is desirable to perform direct connections via reconfigurable qubit interactions.

These considerations make dynamically-reconfigurable hardware platforms such as atomic systems~\cite{bluvstein2024logical,bluvstein2022quantum,ryan-anderson2022implementing,pino2021demonstration} particularly appealing.
In particular, neutral atom arrays have demonstrated hundreds of transversal gate operations on tens of logical qubits, making use of the flexible connectivity afforded by atom moving~\cite{bluvstein2024logical}.
In comparison, while systems with connections based on fixed wiring can support long-range connectivity and switching~\cite{bartolucci2021switch,bravyi2024high}, transversal connections between multiple logical qubits likely increases the cumulative qubit degree which may significantly increase the hardware complexity.
From a clock speed perspective, for typical assumed code distances of $d\sim 30$, our techniques correspond to a 10\,--100$\times$ speed-up by using transversal operations in a reconfigurable architecture.

\end{document}


\title{Supplementary Information: Low-Overhead Transversal Fault Tolerance for Universal Quantum Computation}
\author{Hengyun Zhou}
\thanks{These authors contributed equally}
\email{hyzhou@quera.com}
\affiliation{QuEra Computing Inc., 1284 Soldiers Field Road, Boston, MA, 02135, US}
\affiliation{Department of Physics, Harvard University, Cambridge, Massachusetts 02138, USA}

\author{Chen Zhao}
\thanks{These authors contributed equally}
\affiliation{QuEra Computing Inc., 1284 Soldiers Field Road, Boston, MA, 02135, US}

\author{Madelyn Cain}
\affiliation{Department of Physics, Harvard University, Cambridge, Massachusetts 02138, USA}

\author{Dolev Bluvstein}
\affiliation{Department of Physics, Harvard University, Cambridge, Massachusetts 02138, USA}

\author{Nishad Maskara}
\affiliation{Department of Physics, Harvard University, Cambridge, Massachusetts 02138, USA}

\author{Casey Duckering}
\affiliation{QuEra Computing Inc., 1284 Soldiers Field Road, Boston, MA, 02135, US}

\author{Hong-Ye Hu}
\affiliation{Department of Physics, Harvard University, Cambridge, Massachusetts 02138, USA}

\author{Sheng-Tao Wang}
\affiliation{QuEra Computing Inc., 1284 Soldiers Field Road, Boston, MA, 02135, US}

\author{Aleksander Kubica}
\affiliation{AWS Center for Quantum Computing, Pasadena, California 91125, USA}
\affiliation{California Institute of Technology, Pasadena, California 91125, USA}
\affiliation{Department of Applied Physics, Yale University, New Haven, Connecticut 06511, USA USA}

\author{Mikhail D. Lukin}
\email{lukin@physics.harvard.edu}
\affiliation{Department of Physics, Harvard University, Cambridge, Massachusetts 02138, USA}

\maketitle

\section{Summary of Notation}

To facilitate reading the rest of the supplementary information, we summarize our notation in Tab.~\ref{tab:conventions}.

\begin{table*}
\begin{tabular}{|c|l|}
\hline
$\mathcal{C}$ & Ideal Clifford quantum circuit with magic state inputs and feed-forward operations\\\hline
$|\overline{0}\rangle$ & Logical $|0\rangle$ initial state prepared via random stabilizer projections\\\hline
$|\ooverline{0}\rangle$ & Ideal logical $|0\rangle$ code state, with all stabilizers fixed to +1\\\hline
$\overline{CNOT}$ & Logical CNOT operation\\\hline
$CNOT$ & Physical CNOT operation\\\hline
$M$ & Number of ideal (logical) measurements performed in the ideal (logical) circuit\\\hline
$T$ & Number of gate operation layers\\\hline
$B$ & Maximal number of code blocks at any given time\\\hline
$q$ & Number of logical qubit initializations performed in the Pauli basis\\\hline
$\vec{b}_{\mathcal{C}}\in \mathbb{Z}_2^M$  & Logical bit string sampled from circuit $\mathcal{C}$\\\hline
$\vec{b}_j$ & Vector formed by the first $j$ logical measurement results of a given shot\\\hline
$f_{\mathcal{C}}\in (\mathbb{Z}_2^M\rightarrow \mathbb{R})$ & Distribution of logical bit strings sampled from circuit $\mathcal{C}$\\\hline
$p$ & Parameter characterizing the noise strength\\\hline
$p_{th}$ & Error threshold\\\hline
$\mathcal{Q}$ & Quantum code\\\hline
$r$ & Upper bound on stabilizer weight\\\hline
$c$ & Upper bound on number of stabilizers each qubit is involved in\\\hline
$t$ & Maximal number of qubits within a code block connected by transversal gates \\\hline
$s$ & Size of connected cluster in the decoding hypergraph\\\hline
$v$ & Maximal degree of a node in a hypergraph\\\hline
$d$ & Code distance\\\hline
$\tcal{C}$ & Transversal realization of ideal circuit $\mathcal{C}$\\\hline
$\tcal{C}_e$ & Transversal realization of ideal circuit $\mathcal{C}$ with error realization $e$\\\hline
$\tcal{C}_0$ & Transversal realization of ideal circuit $\mathcal{C}$ with no errors\\\hline
$|_j$ & \makecell{Object for the circuit up to the $j$th logical measurement, e.g. $\tcal{C}|_j$ denotes\\ the transversal realization of the ideal circuit up to the $j$th logical measurement} \\\hline
$\mathcal{E}$ & Set of elementary errors (faults)\\\hline
$e\in \mathbb{Z}_2^{|\mathcal{E}|}$ & A given error realization\\\hline
$\mathcal{D}$ & Set of detectors \\\hline
$\partial e\in \mathbb{Z}_2^{|\mathcal{D}|}$ & Set of detectors a given error $e$ triggers\\\hline
$\Gamma$ & Hypergraph corresponding to the detector error model\\\hline
$\Xi$ & Line graph of $\Gamma$, also known as syndrome adjacency graph in Ref.~\cite{gottesman2013fault}\\\hline
$\kappa\in \mathbb{Z}_2^{|\mathcal{E}|}$ & Recovery returned by the most likely error decoder\\\hline
$\mathcal{G}$ & Set of frame variables, corresponding to distinct patterns of $Z$ operators applied on the $|\ooverline{0}\rangle$ initial state \\\hline
$g\in \mathbb{Z}_2^{|\mathcal{G}|}$ & A given realization of frame variables \\\hline
$g_l$ & Frame logical variable, i.e. a frame variable that commutes with all stabilizers\\\hline
$\Lambda$ & Matrix describing how frame logical variables flip logical measurement results\\\hline
$\lambda$ & An inferred assignment of frame variables that returns the code to the codespace with all stabilizers equal to +1\\\hline
$f=e\oplus\kappa$ & Fault configuration, formed from the mod 2 addition of errors and error recovery operators\\\hline
$h=g\oplus\lambda$ & Frame configuration, formed from the mod 2 addition of frame variables and inferred frame repair operators\\\hline
$P(e)$ & Forward propagation of operator $e$ through the Clifford circuit to logical measurements\\\hline
$z$ & Physical measurement results that a logical measurement corresponds to\\\hline
$z_i$ & \makecell{Physical measurement results that would have occurred\\ if no errors happened after the initial random stabilizer projections}\\\hline
$L_j$ & $j$-th logical operator\\\hline
$\overline{L}(z)$ & Logical measurement result inferred from the physical measurement results $z$\\\hline
$\overline{F}(e)$, $\overline{F}(g)$ & Change in the logical measurement result due to error or frame operators\\\hline
$\overline{L}_c(z,\kappa,\lambda)$ & Corrected logical measurement result after applying the inferred recovery operator $\kappa$ and frame operator $\lambda$\\\hline
\end{tabular}\\
\caption{Summary of conventions employed in this paper.
For the error and frame variables, we use the same notation for both the binary variables and the Pauli operators they correspond to, where the meaning should be clear based on the context.
An overline distinguishes operations and variables at the logical level from the corresponding ones at the physical level.
\label{tab:conventions}}
\end{table*}

\section{Detailed Description of Protocol}
\label{sec:formal_description}

In this section, we provide detailed descriptions of our protocol, further elaborating on the ``key concepts'' section from Methods.
QEC experts may wish to skip ahead to Sec.~\ref{si_sec:frame}, in which we provide a formal description of the randomness associated with measurement-based logical qubit initialization, and clarify how to interpret non-deterministic logical measurement outcomes.

\subsection{Ideal Quantum Circuits}
First, let us describe our protocol for turning a target quantum circuit into a fault-tolerant circuit.
We assume that the circuit is specified in a computational model with Pauli basis state preparation and measurement, single- and two-qubit Clifford gate operations, and $|T\rangle=T|+\rangle$ magic state inputs.

\begin{definition}[Ideal quantum circuit]
\label{def:ideal_circuit}
Define $\mathcal{C}$, an ideal Clifford quantum circuit with magic state inputs and feed-forward operations (henceforth ideal quantum circuit), to be a quantum circuit that consists of layers of the following operations:
\begin{enumerate}
\item Qubit initialization in state $|0\rangle$.
\item Single-qubit $Z$ gates.
\item Single-qubit $H$ gates.
\item Single-qubit $S$ gates.
\item $CNOT$ gate between any pair of qubits.
\item Identity gate, if no other operation is specified on a given qubit.
\item Measurement of a subset of qubits in the $Z$ basis.
\item Feed-forward Clifford operations of the above types. Conditional on certain qubit measurement results, perform some combination of the preceding operations on the remaining qubits.
\item Qubit initialization in the magic state $|T\rangle=T|+\rangle$.
\end{enumerate}
\end{definition}
Note that to simplify the construction of an error-corrected version of these circuits, we have compiled the Clifford circuit into a particular set of operations.
$X$ or $Y$ basis operations can be obtained from the $Z$ basis via $H$ and/or $S$ gates.

\subsection{Noise Models}
In practice, quantum circuits will experience noise.
For our theoretical analysis, we adopt the local stochastic noise model as a simplified description of actual noise channels~\cite{gottesman2013fault}.
Consider a set of possible elementary errors (faults) $\mathcal{E}$, and denote a given error realization by the vector $e\in \mathbb{Z}_2^{|\mathcal{E}|}$, where the $i$th entry of the vector is equal to one if and only if the $i$th error in $\mathcal{E}$ occurred.
The local stochastic noise model satisfies the following property: the probability that an error $e$ of weight $s$ occurs is at most $p^{s}$, where $p$ is the error rate.
For the set of possible elementary errors, we choose the following data-syndrome error set~\cite{gottesman2013fault}: data qubits experience error rate $p$ per initialization, syndrome extraction, and measurement, and the syndrome bit readout experiences error rate $p$.
Following Ref.~\cite{gottesman2013fault}, we do not add extra errors for transversal gates themselves but only the round of syndrome extraction that follow them.
Incorporating gate errors just corresponds to a rescaling of the error rate.
While this error model is simplified compared to experimental noise models, threshold proofs for the former can be readily generalized to the latter by choosing a different set of elementary errors and noting that syndrome extraction circuits for QLDPC codes typically have bounded depth, and therefore error propagation is also bounded.
The change in error model only results in a quantitative modification of the threshold, without changing the overall conclusions.
Thus, we use the simplified local stochastic noise model for our proofs, and more detailed circuit-level noise models (Sec.~\ref{si:numerics}) for numerical simulations.

To capture the effect of a given set of errors on logical measurements, we also define the forward-propagated error $P(e)$.
An error $E$ on some data qubits occurring before a unitary $U$ is equivalent to $UEU^\dagger$ occurring after the unitary.
We can thus propagate any error event forward in time through the circuit.
Note that syndrome measurement errors do not directly act on a physical data qubit, and therefore are not propagated forward.
For a set of errors $e$, we can keep propagating the error forward until it reaches either a logical measurement or an unmeasured output logical qubit of the circuit.
We denote the resulting operator as $P(e)$, and its restriction onto the $j$th logical measurement as $P(e)|_j$.

\subsection{Error-Corrected Quantum Circuits}
We now describe how to realize the ideal quantum circuit in this noisy setting, using logical qubits and quantum error correction.
We consider CSS stabilizer quantum codes $\mathcal{Q}$, encoding $k$ logical qubits into $n$ physical data qubits, with code distance $d$, denoted by the notation $[[n,k,d]]$.
We restrict our attention to quantum low-density parity check (QLDPC) codes, where each stabilizer generator has weight $\leq r$ and each data qubit is involved in $\leq c$ stabilizer generators.
QLDPC codes have the nice property that the resulting syndrome adjacency graph (see following discussion) has bounded degree, thereby causing fault configurations to form small connected clusters that are more easily corrected.
There are many QLDPC code constructions, including surface codes~\cite{kitaev2000unpaired,fowler2012surface}, color codes~\cite{bombin2006topological}, and various high-rate constructions based on products and/or polynomials~\cite{tillich2014quantum,breuckmann2021quantum,bravyi2013homological,hastings2021fiber,breuckmann2020balanced,panteleev2022quantum,panteleev2022asymptotically,kovalev2013quantum,bravyi2024high}.
We only consider the case where all code blocks belong to the same code family, instead of the more general case where different codes may be mixed and matched.
We will also focus on the case where all code blocks have the same size.

Our analysis focuses on transversal operations, which have well-behaved error propagation.
Transversal gates are defined relative to a partition of the code blocks~\cite{eastin2009restrictions,jochym-oconnor2018disjointness}.
We choose the same, fixed partition for all code blocks, and use the parameter $t$ to denote the maximal size of any part within a code block.
We call a physical implementation $U$ of a logical operation $\overline{U}$ transversal with respect to this partition, if it exclusively couples qubits within the same part (see Methods and below for specific examples in state preparation and measurements, as well as gate operations).
We will also focus only on transversal operations consisting of depth-one quantum circuits (excluding SE rounds), which cover most common transversal Clifford gates.
The advantage of transversal gates is that the spread of errors is constrained to be within each partition.

We now consider how a given ideal circuit $\mathcal{C}$ can be implemented using error-correcting codes and transversal operations, subject to the local stochastic noise model described above.

\begin{definition}[Transversal realization $\tcal{C}$ of ideal circuit $\mathcal{C}$]
\label{def:error_corrected_circuit}
Consider an ideal quantum circuit $\mathcal{C}$, and a QEC code $\mathcal{Q}$ with some set of transversal operations.
If there exists a sequence of transversal operations of $\mathcal{Q}$, such that the logical operations implement the ideal quantum circuit on some of the logical qubits, then we call the following circuit a transversal realization $\tcal{C}$ of the ideal circuit $\mathcal{C}$:
\begin{enumerate}
\item Qubit initialization in the $Z$ basis is replaced by initialization of all physical qubits in a code block in $\ket{0}$, followed by one syndrome extraction (SE) round.
\item Single-qubit Pauli gates do not lead to any physical action, but are tracked in the logical Pauli frame.
\item Clifford gate operations are performed via transversal gate operations, including transversal CNOT gates between blocks and fold-transversal gates within each block~\cite{kubica2015unfolding,moussa2016transversal,breuckmann2024fold,quintavalle2023partitioning}, followed by one SE round.
\item Qubit measurement in the $Z$ basis is replaced by measuring all data qubits in a code block in the $Z$ basis.
\item Feed-forward operations and magic state teleportation are executed in the same way as the ideal circuit, based on the decoded logical measurement results.
\item Magic states are assumed to be provided with all stabilizers fixed to +1, followed by local stochastic data qubit noise of strength $p$.
\end{enumerate}
\end{definition}

The magic states are assumed to be prepared via some separate procedure in this formulation.
In practice, as they are often obtained via magic state distillation~\cite{bravyi2005universal} involving Clifford circuits and noisy state injection, we conjecture that our conclusions can be generalized to include these procedures as well.

A simple example of a transversal realization of a circuit is the preparation and measurement of correlations of a Bell pair.
Using the transversal CNOT for CSS codes, we can implement this using two blocks of any CSS QLDPC code, where only one logical qubit in each block is used to create the Bell pair.
This then enables implementing the target circuit with a family of codes with growing distance, allowing us to use the threshold theorem below and achieve exponential error suppression for the given circuit.

Note that for a general CSS QLDPC code family, the above prescription may only allow the implementation of a subset of ideal quantum circuits.
For a quantum code encoding many logical qubits, all logical qubits within a given code block must be initialized and measured in the same basis.
Moreover, transversal gate operations for this code may only be able to implement a subset of Clifford gates~\cite{breuckmann2024fold}.

\begin{figure}
\centering
\includegraphics[width=\columnwidth]{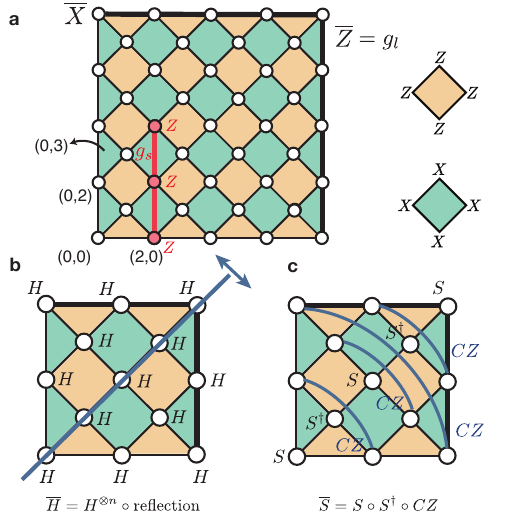}
\caption{(a) Illustration of the non-rotated surface code. White circles indicate data qubits. Orange (green) plaquettes are $Z$ ($X$) stabilizers. The logical $\overline{Z}$ ($\overline{X}$) operator runs vertically (horizontally), and we choose our convention for fixing $Z$ ($X$) stabilizers to be performing a chain of $X$ ($Z$) flips to the left (bottom) boundary, as illustrated by the red line.
We refer to the rows (columns) that have data qubits on the outer edge as major rows(columns).
We have also labeled the qubit coordinate system convention.
(b) Illustration of transversal $\overline{H}$ gate, consisting of physical $H$ gates followed by a reflection along the diagonal.
We choose to perform a reflection instead of rotation to limit the transversal partition size to two.
(c) Illustration of transversal $\overline{S}$ gate, consisting of $S$ and $S^\dagger$ gates along the diagonal, together with $CZ$ gates between mirrored qubits.
}
\label{fig:surface_code}
\end{figure}

However, using the surface code, which has a transversal implementation of the whole Clifford group, we can obtain a transversal realization $\tcal{C}$ of any ideal circuit $\mathcal{C}$.
The same conclusion also applies to other codes with transversal Clifford operations, such as the 2D color code.

We now review the definition of the surface code.
We focus on the non-rotated surface code for our proof, due to the relative simplicity of gate implementations, but we expect the conclusions to readily apply to other variations as well.
We illustrate the non-rotated surface code in Fig.~\ref{fig:surface_code}.
The distance $d$ surface code consists of $n=d^2+(d-1)^2$ data qubits and $n-1$ stabilizers.
The logical operators $\overline{X}$ and $\overline{Z}$ are shown in Fig.~\ref{fig:surface_code} as well.

We can now obtain a transversal realization of any ideal quantum circuit described in Def.~\ref{def:ideal_circuit}.

\begin{definition}[Surface code transversal realization]
\label{def:surface_code_circuit}
Given an ideal quantum circuit $\mathcal{C}$ with magic state inputs and feed-forward operations (Def.~\ref{def:ideal_circuit}), we define its surface code transversal realization $\tcal{C}$ of distance $d$ by replacing each of the operations as follows:
\begin{enumerate}
\item Qubit initialization in the $Z$ basis is replaced by initialization of physical data qubits in $|0\rangle$, followed by one SE round.
\item Single-qubit $Z$ gates do not lead to any physical action, but are tracked in the logical Pauli frame.
\item Single-qubit $H$ gates are replaced by a transversal $H$ gate, in which we apply an $H$ gate on each physical qubit of the code block, followed by a reflection across the diagonal, and one SE round (Fig.~\ref{fig:surface_code}(b)).
\item Single-qubit $S$ gates are replaced by a fold-transversal $S$ gate~\cite{kubica2015unfolding,moussa2016transversal}, in which physical $S$, $S^\dagger$ gates are applied in an alternating fashion on qubits on the diagonal, and $CZ$ gates are applied on pairs of qubits that are matched together when folding across a diagonal (Fig.~\ref{fig:surface_code}(c)). This is followed by one SE round.
\item $CNOT$ gates are replaced by transversal $CNOT$s between pairs of logical qubits, followed by one SE round.
\item Identity gates are replaced by one SE round.
\item Measurements in the $Z$ basis are replaced by a transversal measurement of all corresponding physical qubits in the $Z$ basis.
\item Feedforward Clifford operations are executed in the same way as above, based on the decoded logical measurement results.
\item Magic states are assumed to be provided with all stabilizers fixed to +1, followed by local stochastic data qubit noise of strength $p$.
\end{enumerate}
Here, all logical qubits (code blocks) are non-rotated surface codes of the same code distance $d$.
\end{definition}

The syndrome measurement for the surface code can be performed simultaneously in both bases~\cite{fowler2012surface}.
When initializing the logical qubit, the values in one basis are already deterministic, and therefore we only need to measure the complementary basis.
However, for simplicity of analysis, we include both bases here.

Each logical operation is followed by one SE round in our construction.
This is primarily for simplicity of our analysis, and the number of rounds should be optimized in practice depending on the given target circuit and target logical error rate, possibly even performing multiple gate operations before one SE round~\cite{cain2024correlated}.
Notice also that we never perform $d$ SE rounds following any given operation.

\subsection{Error Correction and Decoding}
\label{si:decoding}
Having specified the error-corrected quantum circuit, let us now describe how we handle errors and interpret logical measurement results.
To start with, we consider the standard decoding approach, in which a detector error model (decoding hypergraph) is constructed, and a recovery operator $\kappa$ is identified that reproduces the observed syndrome patterns.

As above, consider a given error realization $e\in \mathbb{Z}_2^{|\mathcal{E}|}$, where the $i$th entry of the vector is one iff the $i$th error in $\mathcal{E}$ occurred.
We will also use the same notation to denote the Pauli operator the error realization corresponds to, where the meaning should be clear from the context.
In the absence of errors, certain products of stabilizer measurement outcomes are deterministic.
For example, with an idling logical qubit, the product of successive stabilizer outcomes is deterministic in the absence of errors.
We denote these deterministic products as detectors (checks), and a generating set of detectors is denoted as $\mathcal{D}$.
In the presence of an error $e$, some set of detectors will be triggered, which we denote as $\partial e$.
We can construct the detector error model (decoding hypergraph) $\Gamma$, in which vertices are detectors, and (hyper)edges are error events.
This also motivates the boundary operator notation $\partial$, as the boundary of the hyperedges are the detector nodes.
To analyze error clusters, we also introduce the related notion of the syndrome adjacency graph $\Xi$~\cite{gottesman2013fault}.
In this hypergraph, vertices are elementary fault locations, and hyperedges are detectors connecting the fault locations they flip.

Due to the feed-forward operations, the circuit must be constructed in a sequential manner, where the actual circuit to be executed only becomes available after interpreting and committing to past measurement results.
Generically, the circuit $\tcal{C}|_j$ for the $j$th logical measurement is only constructed after interpreting the first $j-1$ logical measurements, and performing any requisite feed-forward operations.
It also varies between different shots of executing the logical algorithm, due to randomness in the measurement results.
Similarly, we construct a decoding problem $\Gamma|_j$ for each shot based on the circuit and errors that occur up to the $j$th logical measurement.
In the following, we will use $\Gamma|_j$ to determine the $j$th logical measurement.
We also ensure that the assigned measurement results for the first $j-1$ logical measurements are consistent with the feed-forwards and circuits chosen, as discussed below.

Let us now discuss the concrete construction of detectors for the surface code.
The construction can be readily extended to the case of general LDPC codes.
We construct detectors in a time-local fashion, using the fact that logical gate operations are interspersed with SE rounds.
To describe the detectors, we label the syndrome extraction result with the logical qubit index $i$, syndrome round index $r$, location within code block $(x,y)$, and basis $B=X$ or $Z$.
Our physical qubit location coordinate system starts from the bottom left, with the bottom left data qubit having the label $(0,0)$.
We place data qubits at coordinates $(x,y)$ with $x+y\equiv 0\mod 2$, e.g. the next data qubit to the right is at coordinate $(2,0)$ (Fig.~\ref{fig:surface_code}(a)).
Stabilizers are placed at the center of the corresponding plaquette.
With this convention, we can label the measurement result of the bottom left $Z$ stabilizer of logical qubit 1, in round 3, as $S(i=1,r=3,x=1,y=0,B=Z)$.
The first stabilizer measurement round is labeled round 1.
For initialization in the $\overline{Z}$ basis, we set the round 0 $Z$ stabilizer values to be +1, since they are initialized with a deterministic eigenvalue, and construct a detector comparing the round 1 $Z$ stabilizer value with this.
Meanwhile, the $X$ stabilizer values are random and hence there is no detector comparing the first $X$ stabilizer value to previous results.
For measurements in the $\overline{Z}$ basis, we construct a final round $Z$ stabilizer value by multiplying the measurement results of the corresponding data qubits, and do not assign $X$ stabilizer values since they are unknown when performing a transversal $Z$ measurement.
The detectors can now be constructed for each of the logical operations as follows:

\begin{enumerate}
\item For an identity gate on logical qubit $i$ before SE round $r$, the detector is
\begin{align*}
S(i,r-1,x,y,B)S(i,r,x,y,B).
\end{align*}
\item For a $\overline{H}$ gate on logical qubit $i$ before SE round $r$, the detectors are
\begin{align*}
S(i,r-1,x,y,X)S(i,r,y,x,Z),\\
S(i,r-1,x,y,Z)S(i,r,y,x,X).
\end{align*}
In other words, we compare against the stabilizer after mirroring across the diagonal.
\item For a transversal $\overline{CNOT}$ from logical qubit $i$ to logical qubit $j$, before SE round $r$, the detectors are
\begin{gather*}
S(j,r-1,x,y,X)S(j,r,x,y,X),\\
S(i,r-1,x,y,Z)S(i,r,x,y,Z),\\
S(i,r-1,x,y,X)S(j,r-1,x,y,X)S(i,r,x,y,X),\\
S(i,r-1,x,y,Z)S(j,r-1,x,y,Z)S(j,r,x,y,Z).
\end{gather*}
See also Ref.~\cite{cain2024correlated}. The transversal $\overline{CNOT}$ propagates $X$ errors from control to target, and $Z$ errors from target to control, thereby leading to the higher-weight detectors.
\item For a $\overline{S}$ gate on logical qubit $i$ before SE round $r$, the detectors are
\begin{gather*}
S(i,r-1,x,y,Z)S(i,r,x,y,Z),\\
S(i,r-1,x,y,X)S(i,r-1,y,x,Z)S(i,r,x,y,X).
\end{gather*}
This bears some similarity to the $\overline{CNOT}$ gate, but couples the $X$ and $Z$ components of the decoding problem together rather than that of two logical qubits. More precisely, the measurement error of the $Z$-stabilizer on the control qubit at position $(r-1, x, y)$ before $\overline{CNOT}$ is analogous to the measurement error of the $Z$-stabilizer at position $(r-1, x, y)$ before $\overline{S}$, because they both flip two detectors in round $r$, resulting in a weight-3 hyperedge when combined with previous rounds. Similarly, the measurement error of the $Z$-stabilizer on the target qubit at position $(r-1, x, y)$ before $\overline{CNOT}$ is analogous to the measurement error of the $X$-stabilizer at position $(r-1, x, y)$ before $\overline{S}$, because they only flip one single detector in round $r$.
\end{enumerate}

Given the detector error model and a detector shot $\partial e$, a decoder returns a recovery operator $\kappa$, such that $\partial\kappa=\partial e$.
The total action of error and recovery is then given by $f=e\oplus\kappa$, where addition is understood to be mod 2.
In slight abuse of terminology, we will refer to this joint action as the fault configuration.
By linearity, we have that $\partial(\kappa\oplus e)=0$.
For the purposes of our discussion, we will make use of the most likely error (MLE) decoder, also known as the minimum weight decoder.
The MLE decoder returns the \textit{most likely error} $\kappa\in \mathbb{Z}_2^{|\mathcal{E}|}$ that is consistent with the observed detectors.
Note that this decoder solves the most likely error problem instead of the maximum likelihood problem, i.e. it does not consider the entropy factor associated with the number of cosets.
Additionally, for generic decoding problems, identifying the most likely error may be computationally challenging, although efficient heuristics exist (see Decoding Complexity, Methods).

\subsection{Logical Qubit Initialization and Frame Variables}
\label{si_sec:frame}

We now introduce some useful concepts to describe the randomness associated with measurement-based logical qubit initialization, and clarify how to interpret random logical measurement outcomes.

Due to the random initial projection when measuring $X$ stabilizers during $\lket{0}$ initialization, the physical state is not initially in the code space, where all stabilizers should have eigenvalue +1.
To describe this, we adapt and formalize a concept introduced and implemented in Stim~\cite{gidney2021stim}.
There, to capture the randomness introduced when measuring a physical qubit initialized in $|0\rangle$ in the $X$ basis, a $Z$ operator on that site is multiplied into the state with 50$\%$ probability.
Starting from a reference sample of measurement results, the full measurement result distribution can then be obtained by considering the distribution over these random $Z$ operators and error events.
We refer to these $Z$ operators that act on the initialization boundary as ``frame operators" ($Z$ operator acting on each physical qubit of the initial logical qubit), and variables describing them as ``frame variables", where the name is meant to indicate that they describe the reference frame of random stabilizer initialization, and the reference frame in which we will interpret our logical measurement results.

Formally, consider a $\lket{0}$ logical qubit initialization.
For each data qubit in the code block, associate a $Z$ operator.
Some of these $Z$ operators will have inter-dependencies due to $Z$ stabilizer constraints.
Therefore, we can construct a basis $\mathcal{G}$ of frame operators as follows:
For each data qubit $i$, we add $Z_i$ to $\mathcal{G}$ if it is not equivalent to any combination of operators in $\mathcal{G}$ up to stabilizers.
For an $[[n,k,d]]$ quantum code with $r_Z$ independent $Z$ stabilizer generators, we have $|\mathcal{G}|=n-r_Z$.
We use $g$ to denote both a product of frame operators taken from $\mathcal{G}$ and a binary vector $g\in \mathbb{Z}_2^{|\mathcal{G}|}$ describing it.

Some of the frame operators will flip $X$ stabilizers, and correspond to different effective code spaces (reference frames) that we may project into during the initial random stabilizer measurement results.
We denote these by $g_s$, and refer to them as frame stabilizer operators.
There are also frame operators $g_l$ that do not flip any $X$ stabilizers, instead corresponding to a logical $\overline{Z}$ operator of the code block.
We refer to them as frame logical operators.
While applying these frame logical operators does not change the initial physical state $\lket{0}$, it does lead to different interpretations of the logical measurement result without changing the measurement distribution, a fact that is crucial for our construction.
To capture the relation between frame logical operators $g_l$ and logical measurement results that they might flip, for each circuit $\tcal{C}|_j$, we introduce a matrix $\Lambda\in \mathbb{Z}_2^{j\times q_j}$, where $q_j$ is the number of logical initializations in the Pauli basis (thereby producing $q_j$ frame logical operators), and $j$ is the number of logical measurements that have been performed up to this point.
Note that if more than one logical qubit is encoded in each code block, there will be as many $Z$ frame logical operators as there are logical qubits.
For a given circuit $\tcal{C}|_j$, $\Lambda$ can be efficiently constructed by propagating the frame logical operators until they reach the logical measurements, using standard techniques for propagating Pauli operators through Clifford circuits.

As a concrete example, let us define a basis of frame operators for the surface code (Fig.~\ref{fig:surface_code}).
As mentioned above, we will choose $\overline{X}$ to be the product of $X$ operators on the top row, and $\overline{Z}$ to be the product of $Z$ operators on the rightmost column.
We choose the frame logical operator to be the $\overline{Z}$ logical operator representative above.
For each $X$ stabilizer $s$, we choose a frame operator $g_s$ consisting of a string of $Z$ operators along the column that the $X$ stabilizer is located in, starting from the bottom data qubit of the stabilizer and ending at the bottom boundary (see red line in Fig.~\ref{fig:surface_code}(a)).
By definition, $g_s$ will only flip the single stabilizer $s$, while all other stabilizers and logical operators remain unchanged.
Together, these form a basis $\mathcal{G}$ of frame operators for the surface code.
While any equivalent choice of logical qubit and frame operators is valid (Lemma~\ref{lemma:frame_trivial}), we choose this particular convention so that fixing the stabilizer values will not change the logical qubit readout result.

\subsection{Interpreting Logical Measurement Results}

With these concepts in hand, we now consider how logical measurement results are interpreted, particularly in the case where the logical measurement results are random.
The majority of error correction analyses and simulations focus on the case of a deterministic observable, as they provide a simple characterization of logical error rates.
However, the case of non-deterministic observables is equally important, and the interpretation of them can be more intricate.

To start with, let us describe the logical qubit initialization procedure in terms of frame variables.
To initialize a logical qubit in $\lket{0}$, we start with all physical qubits in $\ket{0}$, and perform a single SE round.
This projects the $X$ stabilizers to take on random values.
The quantum state can be described in terms of frame variables as $\lket{0}=g|\ooverline{0}\rangle$, where $|\ooverline{0}\rangle$ is the ideal $\ket{0}$ logical state with all stabilizers fixed to +1, and $g$ is some appropriate frame variable.
Intuitively, we start from $|\ooverline{0}\rangle$ and flip certain $X$ stabilizers to reach the actual state $\lket{0}$.
Similar to the error variable $e$, the frame variable $g$ will not be directly accessible to us, and must be inferred from our observations.

We will now describe the procedure of interpreting the logical measurement outcome of a noisy error-corrected quantum circuit in three steps.

First, we apply the standard decoding procedure in Sec.~\ref{si:decoding} to obtain an inferred error recovery operator $\kappa$, such that $\partial(e\oplus\kappa)=0$.
This ensures that the resulting frame configuration $f=e\oplus\kappa$ does not trigger any detectors.

Next, we perform the analogous procedure to error recovery for the frame variables, which we refer to as a frame repair operation $\lambda\in\mathbb{Z}_2^{|\mathcal{G}|}$.
Whereas error recovery aims to ensure that no detectors are triggered in the bulk of the quantum circuit, frame repair aims to ensure that we return to the ideal code space with all stabilizers set to +1 when interpreting a logical measurement.
Therefore, we choose $\lambda$, such that the combined effect of error operator $e$, recovery operator $\kappa$, frame operator $g$ and frame repair operator $\lambda$ does not violate any stabilizers.
In other words, $(e\oplus\kappa)\oplus(g\oplus\lambda)$ should act as a stabilizer or logical operator.
We refer to the combination $h=g\oplus\lambda$ as the ``frame configuration", again mirroring the notation for faults.

Finally, we evaluate the logical observable after applying the above corrections.
Denote the raw logical observable inferred from the bit strings as $$\overline{L}(z)=\bigoplus_{z_i\in \overline{L}} z_i.$$

The raw logical observable already incorporates the effect of $e$ and $g$, which physically occurred.
To obtain the corrected logical observable, we propagate the effects of the error recovery operation $\kappa$ and frame repair operation $\lambda$ to the measured logical qubit.
Recalling that $P(\kappa)|_j$ denotes the forward propagation of operator $\kappa$ to the $j$th logical measurement $\overline{L}$, we can define the parity flip of the logical observable due to $\kappa$:
\begin{align}
\overline{F}(\kappa)=
\begin{cases}
0, & \left[P(\kappa)|_j, \overline{L}\right]=0,\\
1, & \left[P(\kappa)|_j, \overline{L}\right]\neq 0,
\end{cases}
\end{align}
where the bracket indicates taking the commutator.
The corrected logical observable is then given by
\begin{align}
\overline{L_c}(z,\kappa,\lambda)=\overline{L}(z)\oplus \overline{F}(\kappa)\oplus \overline{F}(\lambda).
\end{align}

Now let us consider how the error recovery and frame repair procedures affect the logical measurement result.

First, consider the case when the inference reproduces the error and frame operators applied exactly, i.e. $(e\oplus\kappa)\oplus(g\oplus\lambda)$ is the identity operator.
In this case, the circuit and quantum state are equivalent to preparing ideal code states with all stabilizers set to +1, executing the logical circuit, and performing logical measurements.
As everything is ideal and all stabilizers are +1 throughout, standard arguments show that the logical quantum circuit $\tcal{C}$ executes the ideal quantum circuit $\mathcal{C}$ correctly and reproduces the same distribution of logical bit strings $f_{\tcal{C}}=f_{\mathcal{C}}$.

Next, consider the case where no errors were applied, but we still have the initial random stabilizer projection described by the frame operator $g$, and our frame repair operation $\lambda$ may differ from $g$.
Define a fixed transversal Clifford circuit with magic state inputs $\tcal{C}_{fix}$ by taking a transversal realization $\tcal{C}$, fixing the first $j-1$ logical measurement results and their resulting feed-forward operations, and considering the quantum circuit up to the $j$th logical measurement, thereby obtaining a non-adaptive quantum circuit $\tcal{C}_{fix}$.
We can then show the following lemma:

\begin{lemma}[Frame variables do not affect measurement distribution]
\label{lemma:frame_trivial}
Consider a fixed transversal Clifford circuit with magic state inputs $\tcal{C}_{fix}$ and a fixed, arbitrary fault configuration $f=e\oplus\kappa$ such that $\partial f=0$.
Then for any choice of frame configuration $h=g\oplus\lambda\in \mathbb{Z}_2^{|\mathcal{G}|}$, the corrected logical observable $\overline{L_c}$ has the same measurement distribution regardless of the choice of $h$.
\end{lemma}

Consider the difference in the corrected logical observable between $h=g\oplus\lambda$ and $h_0=I$.
By construction, the combination $h=g\oplus\lambda$ must return the logical qubit to the codespace.
$h$ must thus commute with all $X$ stabilizers, and as $h$ is composed of $Z$ operators, it can therefore only be a combination of $Z$ stabilizers and $Z$ logical operators.
By definition, $h$ is applied on the ideal logical initial state $|\ooverline{0}\rangle$, so we conclude that $h$ acts as a logical $\overline{Z}$ operator on $|\ooverline{0}\rangle$, i.e. the corrected logical observable has the same measurement distribution for $h$ and $h_0$.
\qed

Intuitively, this is because what random stabilizer pattern we projected into should not affect the logical measurement results.
It is important to emphasize that this statement only applies to the \textit{distribution} of measurement results: for any given shot, different choices of frame variables may still lead to different interpretations, a feature that we will make use of in our decoding strategy.

Note that this lemma is formulated in the case of a fixed circuit, which will not generally be the case in the presence of feed-forward operations.
In the latter case, we can still make use of this lemma as follows: consider the full conditional circuit $\tcal{C}_{cond}$, the fixed circuit corresponding to the given branch of conditional operations $\tcal{C}_{fix}$, as well as their corresponding ideal versions $\mathcal{C}_{cond}$ and $\mathcal{C}_{fix}$.
Lemma~\ref{lemma:frame_trivial} shows that for a noiseless circuit, the measurement distribution of the ideal and error-corrected circuits are identical, $f_{\tcal{C}_{fix}}=f_{\mathcal{C}_{fix}}$, regardless of the frame variables.
This immediately implies that the marginal distribution conditioned on some fixed set of previous measurement results are identical.
On the other hand, conditioned on the fixed set of previous measurement results, the fixed and conditional circuits are identical, i.e. $f_{\tcal{C}_{cond}}=f_{\tcal{C}_{fix}}$, $f_{\mathcal{C}_{cond}}=f_{\mathcal{C}_{fix}}$.
This implies that $f_{\tcal{C}_{cond}}=f_{\mathcal{C}_{cond}}$.
Thus, Lemma~\ref{lemma:frame_trivial} can be readily applied to the setting with feed-forward operations as well.

Finally, we briefly comment on the case with noisy operations, with more details provided in the proofs in the next section.
In this case, we first apply the error recovery operator $\kappa$, which handles any detectors in the bulk.
As some error clusters may have terminated on the initialization boundary, the total effect of $e\oplus\kappa$ may lead to both logical errors and a change in the reference frame.
We therefore make corresponding modifications to the frame repair operation $\lambda$ as well, before applying the preceding arguments.

\subsection{Decoding Strategy}
In this section, we provide a description of our full decoding strategy, which includes decoding errors, inferring frame variables, and interpreting logical measurement results.
As described in the main text, the key idea is to perform correlated decoding across the logical algorithm, thereby utilizing all relevant syndrome information.
However, we may need to apply additional frame operators in order to ensure that the executed quantum circuit feed-forward is consistent with past logical measurement results.

When executing transversal Clifford quantum circuits, decoding and performing recovery operations are only necessary when interpreting logical measurement results (i.e. the classical outputs of the quantum computation), which can lead to different executed circuits due to feed-forward operations.
To capture this dependency, we sort the set of logical measurements into an ordering $\{\bar{L}_1, \bar{L}_2, \bar{L}_3,...,\bar{L}_M\}$ based on the time they occur and conditional dependencies.
We require that for any pair $i<j$, the logical measurement result $\bar{L}_i$ must not depend on the subsequent logical measurement result $\bar{L}_j$.
If multiple measurements occur simultaneously, then we can place them in any order, since there are no direct inter-dependencies.

We can now recursively define our decoding strategy.
For each logical measurement $\bar{L}_j$, we assume that the previous logical measurement results $\{\bar{L}_1,...,\bar{L}_{j-1}\}$ have been decoded and interpreted, and we have committed to these previous results in order to perform any necessary feed-forward operations.

\begin{definition}[Decoding strategy]
\label{def:decoding_strategy}
For the $j$th logical measurement, we perform two steps to decode and interpret the measurement result:
\begin{enumerate}
\item \textbf{Partial decoding (correlated decoding):} Based on the current circuit $\tcal{C}|_j$ up to the $j$th logical measurement (including the applied feed-forward operations), construct the detector error model $\Gamma|_j$. Apply the MLE decoder to $\Gamma|_j$ and the available detector data $\mathcal{D}|_j$ to identify and apply (in the Pauli frame) an inferred error recovery operator $\kappa$. From this, obtain logical measurement values $\vec{b}_j^{(1)}$ for the first $j$ logical measurements, where the superscript $(1)$ denotes the first step of decoding.
\item \textbf{Consistency check:} Solve the linear equation over $\mathbb{Z}_2$
\begin{align}
\qty(\Lambda g_l)_{1,...,j-1}= \left(\vec{b}_j^{(1)}\right)_{1,...,j-1}\oplus\vec{b}_{j-1},
\label{eq:consistency_equation}
\end{align}
where the outer subscript denotes taking the first $j-1$ components of the vector, and $\vec{b}_{j-1}$ are the first $j-1$ logical measurement results that we have already committed to.
If there is a solution $g_l$, apply the frame logical operators $g_l$ and update the logical measurement result $\vec{b}_j=\vec{b}_j^{(1)}\oplus \Lambda g_l$, committing to the $j$th measurement result (this guarantees consistency with the first $j-1$ logical measurement results).
If not, a heralded failure has occurred and we abort the execution.
\end{enumerate}
\end{definition}

Notice that each time we perform partial decoding, we only commit to the logical measurement result, without committing to the corrections and reference frame throughout.
In other words, we only commit to the minimal amount of information necessary to determine the feed-forward operations.
We leave possible relaxations of this, where more pieces of information are fixed, to future work.

In this definition, we processed the logical measurement results and feedforward one by one.
The technique, however, also readily applies to the case where we instead partition the logical measurements based on layers of Clifford feedforward operations, resulting in fewer rounds of decoding.

To show that this decoding strategy has a high probability of success, we need to show two things: first, the probability of a heralded error should be low; second, the probability of a regular logical error should be low, such that the measurement distribution should be close in total variation distance (TVD) to the measurement distribution of the ideal circuit.
We will now prove these statements.

\section{Proof of Fault Tolerance}
\label{sec:fault_tolerance_proof}

In this section, we prove the fault tolerance of our decoding strategy, in a setting where we assume fast classical decoding and access to low-noise magic states.
Compared to existing analyses of fault tolerance of qLDPC codes~\cite{gottesman2013fault}, there are two main differences: First, since multiple rounds of correlated decoding may lead to differing interpretations of the same logical measurement result (Fig.~3b of main text and Sec.~\ref{si_sec:consistency}), we need to carefully reason about the process of restoring consistency and show that this is possible with high probability; Second, instead of analyzing each gadget separately in a modular fashion, we must analyze the logical algorithm as a whole.
We will first characterize the effect of low weight physical errors within a single decoding round in Sec.~\ref{si_sec:physical_error}, showing that they cannot lead to a flip of logical measurement results.
In Sec.~\ref{si_sec:logical_error}, we then use this to show that any type of logical error---either a logical flip or an inconsistency between multiple decoding rounds---requires $\Theta(d)$ physical errors to occur.
In Sec.~\ref{si_sec:threshold}, we combine this with standard methods for counting the probability of error clusters of a given size to bound the total logical error rate and demonstrate exponential error suppression.

\subsection{Characterizing the Effect of Errors}
\label{si_sec:physical_error}

We start by examining how physical errors propagate under transversal gate operations.
Transversality guarantees that a given error cannot cause too many errors on a given code block when propagated to the qubit measurements.

\begin{lemma}[Transversal gates limit error propagation]
\label{lemma:transversal}
Consider a transversal realization $\tcal{C}$ of an ideal circuit, with maximal size $t$ of the fixed transversal partition.
Then any fault configuration $f$, when forward propagated $P(f)$ to any logical measurement, has support on at most $t|f|$ data qubits, where $|f|$ is the weight of the fault configuration $f$.
\end{lemma}

This lemma is a straightforward consequence of the definition of transversal gates.
By construction, each individual error can only spread to at most $t$ qubits within each code block, and therefore $P(f)$ has support on at most $t|f|$ data qubits on each code block.
\qed

For our data-syndrome noise model, in which errors occur on data qubits between SE rounds and on the syndrome value itself, we do not need to consider error propagation due to the syndrome extraction procedure itself.
For practical SE circuits, for QLDPC codes with bounded syndrome extraction circuit depth, this only produces a constant factor difference due to the bounded error propagation and doesn't change our qualitative conclusions.

As discussed above, for the special case of the non-rotated surface code and the set of transversal operations that we consider, we have $t=2$.
Intuitively, an error can only affect the given qubit and the qubit that it is paired with via a reflection across the diagonal (Fig.~\ref{fig:surface_code}).

We now introduce a technical lemma characterizing the effect of the fault configuration and frame configuration after applying decoding and error correction.
Here and below, we will use the notation with a hat $\hat{\lambda}$ to indicate a frame repair operator that leads to trivial logical action on the logical measurements of interest for the given shot, and that without a hat $\lambda$ to indicate any other frame repair operator, e.g. those determined from consistency checks.

\begin{figure*}
\centering
\includegraphics{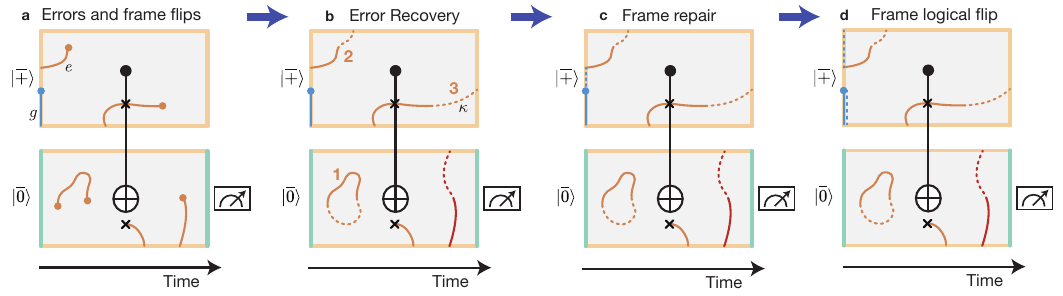}
\caption{\textbf{Illustration of error recovery and frame repair procedures.}
We illustrate the procedure for the surface code, where a cross-sectional view with one spatial axis and one time axis is shown.
We only illustrate $X$ errors and $Z$ stabilizer measurement errors, which are relevant to interpreting the $\overline{Z}$ measurement.
$X$ errors can terminate on orange boundaries, but cannot terminate on cyan boundaries.
The transversal $\overline{CNOT}$ copies $X$ errors from the top to the bottom, resulting in a branching point (black cross) and an error cluster spanning both code blocks.
(a) Error chains and frame flips. Chains of $X$-type errors (orange lines) lead to syndromes (end points) or terminate on appropriate boundaries.
A line segment in the vertical direction is a data qubit $X$ error, while a line segment in the horizontal direction is a measurement error.
Note that the $X$-type error cannot terminate on the transversal $Z$ measurement boundary. The random stabilizer initialization leads to a frame configuration on the logical $\lket{+}$ initialization, as illustrated by the blue line and the flipped $Z$ stabilizer (blue point).
This is similar to the frame stabilizer operator $g_s$ illustrated in Fig.~\ref{fig:surface_code}(a).
(b) We first infer an error recovery operator, which has the same boundary as the error chain.
Together, the error and recovery operator form the fault configuration, which triggers no detectors.
We illustrate a few examples (orange lines) that do not lead to a logical error: (1) the fault configuration forms a closed loop and is equivalent to applying a stabilizer; (2) the fault configuration terminates on an initialization boundary; (3) the fault configuration terminates on an out-going, unmeasured logical qubit, but the forward-propagated errors onto the measured logical qubit are equivalent to a stabilizer.
A logical error can only happen when the fault configuration spans across two opposing spatial boundaries (red line), which requires an error of weight $\Theta(d)$.
(c,d) The frame repair operation returns the logical qubit to the code space with all stabilizers +1, corresponding to cancelling any residual flipped stabilizers on the initialization boundary.
Note that the error recovery process may also lead to a change that needs to be accounted for by frame repair.
An example choice of frame repair is shown in (c), which applies an overall $X$ operator on the logical measurement result.
Alternatively, a different choice of frame repair shown in (d), related to the previous one by a frame logical flip, results in identity operation on the logical measurement result.
}
\label{fig:proof_sketch_si}
\end{figure*}

\begin{lemma}[Correction on codespace for low weight faults]
\label{lemma:trivial_action}
Consider the $j$th logical measurement and the associated syndrome adjacency graph $\Xi|_j$ in a given execution of the transversal realization $\tcal{C}$.
Consider any fault configuration $f=e\oplus \kappa$ in $\Xi|_j$, where the largest weight of any connected cluster of error vertices is less than $d/t$.

Then there exists a choice of frame repair operator $\hat{\lambda}$, such that the combined effect of fault configuration and frame configuration $P(e\oplus \kappa)\oplus P(g\oplus \hat{\lambda})$ does not flip the results of any of the $j$ logical measurements.
\end{lemma}

As we described in Def.~\ref{def:error_corrected_circuit}, measurements are performed in the $Z$ basis ($X$ basis measurements can be performed by an $\overline{H}$ gate followed by a $\overline{Z}$ measurement).
To show that no logical measurement result is flipped, i.e. the combined effect is trivial, we need to show that for the $Z$ measurements we perform, $P(f)\oplus P(g\oplus \hat{\lambda})$ acts as a combination of stabilizers and the logical $\overline{Z}$ operator, which will not change the logical measurement result.
Here, $f=e\oplus \kappa$ as before.

We will analyze each connected cluster $f_i$ of $f$ separately.
Because the different clusters are disjoint, $\partial f=0$ implies that $\partial f_i=0$.
As $P(f)$ is linear in the input error, we can analyze the effect of each connected component independently.
By Lemma~\ref{lemma:transversal} and the condition that $\textrm{wt}(f_i)<d/t$ for all $i$, we have that $\textrm{wt}(P(f_i))<d$.

We now show that $\partial f_i=0$ and $\textrm{wt}(P(f_i))<d$ implies that there exists a choice of frame repair operator $\hat{\lambda}_i$, such that $P(f_i)\oplus P(g\oplus \hat{\lambda}_i)$ acts trivially on the logical measurements.
If this is the case, then we can combine the fault configurations
\begin{align}
f=\bigoplus_i f_i,\label{eq:combine_fault}  
\end{align}
and combine the frame configurations 
\begin{align}
\hat{\lambda}=g\oplus\qty(\bigoplus_i (\hat{\lambda}_i\oplus g)).\label{eq:combine_frame}
\end{align}
Since each component $P(f_i \oplus (\hat{\lambda}_i\oplus g))$ has trivial logical action on the measurement results, by linearity, so does the combined effect $P(f\oplus (g\oplus\hat{\lambda}))$.

Case 1: First, we consider the case where $f_i$ is not connected to any detectors that involve an initial syndrome measurement during $\lket{0}$ state initialization.
Intuitively, this is the case where $f_i$ is not connected to the initialization boundary.
In this case, if we choose $\hat{\lambda}_i=g$, then the combined action of fault configuration and frame configuration is $P(f_i)\oplus P(g\oplus \hat{\lambda}_i)=P(f_i)$.
Since $\partial f_i=0$ and $f_i$ is not connected to an initialization boundary, $P(f_i)$ must be a product of $X$ stabilizers and $X$ logical operators.
Since $\textrm{wt}(P(f_i))<d$, it cannot be a logical $\overline{X}$ operator that changes the $\overline{Z}$ measurement result.
Therefore, it does not flip the logical measurement result.

Case 2: Now consider the case where $f_i$ is connected to a $\lket{0}$ initialization boundary.
Let $\lambda_i$ be the frame repair operation we choose, and let $h_i=g\oplus\lambda_i$.
If $P(f_i\oplus h_i)$ does not flip the logical measurement result, then we have already satisfied our requirements, and we can set $\hat{\lambda}_i=\lambda_i$.
Otherwise, suppose $P(f_i\oplus h_i)=\overline{L}$, where $\overline{L}$ is some nontrivial logical operator.
Since $\textrm{wt}(P(f_i))<d$, the logical operator must have some contribution from the frame configuration $h_i$ located on the initialization boundary.
Similar to Eq.~(\ref{eq:consistency_equation}), we can thus find a frame logical operator $g_l$ to apply on the initialization boundary, such that $P(g_l)=\overline{L}$.
Choosing $\hat{\lambda}_i=\lambda_i\oplus g_l$ then implies that $P(f_i\oplus(g\oplus \hat{\lambda}_i))$ cancels the application of $\overline{L}$ on the logical measurement, such that $\hat{\lambda}_i$ acts trivially on the logical measurements.

Combining the fault configurations and frame configurations as in Eqs.~(\ref{eq:combine_fault},\ref{eq:combine_frame}), the frame configuration $\hat{\lambda}$ will be such that the combined effect of fault configuration and frame configuration does not flip any of the $j$ logical measurements.
\qed

When considering clusters connected to initialization boundaries, we only need to consider those connected to Pauli basis initialization boundaries and not $\lket{T}$ magic state inputs.
This is because per Def.~\ref{def:error_corrected_circuit}, the magic states are provided with known stabilizer values up to local stochastic noise.
As the stabilizer values are known with confidence, detectors can be constructed in both bases to detect and correct any errors nearby.
In other words, unlike $\lket{0}$ initialization, errors cannot terminate on magic state inputs without being detectable.

Lemma~\ref{lemma:trivial_action} only requires such a frame repair operation to exist, but does not require us to explicitly apply it.
We simply use its existence to guarantee consistency between multiple rounds of decoding.
The reason that finding this particular frame repair operation is not important is that per Lemma~\ref{lemma:frame_trivial}, this choice does not affect the logical measurement distribution.
Our decoding strategy only requires us to find frame repair operations that guarantee consistency between multiple rounds of decoding of the same logical measurement in a given shot, without requiring the logical action for a given shot to be trivial.

Let us briefly illustrate this lemma in the case of the surface code.
In Fig.~\ref{fig:proof_sketch_si}(a), we illustrate an instance of physical errors $e$ (orange lines) and initial random frame projection $g$ (blue line).
The error recovery and frame repair procedures are illustrated in Fig.~\ref{fig:proof_sketch_si}(b,c), canceling any bulk detectors and returning the stabilizers to the +1 subspace, respectively.
We illustrate different types of clusters that can appear in Fig.~\ref{fig:proof_sketch_si}(b): case 1 in Lemma~\ref{lemma:trivial_action} is illustrated with the orange lines labeled 1 and 3, while case 2 is illustrated by the orange line labeled 2.
For case 1, the frame configuration acts trivially when forward-propagated to the logical measurement, automatically satisfying our requirements.
For case 2, the frame configuration flips some stabilizers, which we take into account in the frame repair stage (Fig.~\ref{fig:proof_sketch_si}(c)).
After the error recovery and frame repair stage, it is possible that an overall $X$ operator is applied at initialization, which propagates through the CNOT to flip the logical measurement result.
However, one can choose an alternative frame configuration (Fig.~\ref{fig:proof_sketch_si}(d)), which negates the logical operator and thereby acts trivially on the logical measurement, as required by Lemma~\ref{lemma:trivial_action}.

\subsection{Characterizing Logical Errors}
\label{si_sec:logical_error}

It is important to note that we are only guaranteed to not flip the logical qubits that we have performed a measurement on, and only in the basis that we measured.
There could be residual errors on the remaining qubits, or a $Z$ flip on a logical qubit measured in the $Z$ basis.
However, the former will get fixed in later rounds of decoding, so long as we can maintain consistency on the logical measurement results, while the latter does not influence any measurement results.
Thus, they should not cause any effects on the logical measurement distribution.
We formalize this idea in the following key lemma, which characterizes the structure of logical errors.
It shows that small clusters of errors cannot give rise to logical errors on logical qubits that have been measured.

\begin{lemma}[Logical errors must be composed of at least $d/t$ faults]
\label{lemma:logical_error_weight}

Consider the $j$th logical measurement and the associated syndrome adjacency graph $\Xi|_j$ in a given execution of the transversal realization $\tcal{C}$.
Consider any fault configuration $f=e\oplus \kappa$ in $\Xi|_j$, where the largest weight of any connected cluster of error vertices is less than $d/t$.

Then there exists a choice of frame repair operator $\lambda_j$, such that
\begin{enumerate}
\item The first $j-1$ measurement results are consistent with the previous rounds of decoding, if the previous rounds of decoding also satisfy the same conditions above.
\item The distribution of the $j$th measurement, conditioned on the outcome of the first $j-1$ measurement results from the previous round of decoding, is identical to the ideal distribution.
\end{enumerate}
\end{lemma}

Recall our notation convention, where $\hat{\lambda}$ indicates a frame repair operator that has trivial action for the given shot, and $\lambda$ is the frame repair operator that we apply based on consistency conditions.
In other words, \mbox{$P(f)\oplus P(\hat{h})$} has trivial logical action when the latter term $\hat{h}=g\oplus\hat{\lambda}$ has a hat.

Let us start by proving property 2.
The proof here is similar to our discussion following Lemma~\ref{lemma:frame_trivial}.
Conditioned on the outcome of the first $j-1$ measurement results, the circuit up to the $j$th measurement is now a deterministic circuit $\tcal{C}_{fix}$, and we can construct a given syndrome adjacency graph $\Xi|_j$.
By Lemma~\ref{lemma:trivial_action}, there exists a choice of frame repair operator $\hat{\lambda}$ that produces the same measurement distribution as the corresponding fixed ideal circuit $\mathcal{C}_{fix}$, i.e. $f_{\tcal{C}_{fix}}=f_{\mathcal{C}_{fix}}$.
In particular, conditioned on the first $j-1$ measurement results, it also reproduces the marginal distribution of the $j$th measurement result.

Conditioned on the first $j-1$ measurement results, the fixed circuit $\mathcal{C}_{fix}$ and adaptive circuit $\mathcal{C}$ are identical and have the same marginal distribution for the $j$th logical measurement, for both the ideal and error-corrected case, i.e. $f_{\mathcal{C}}=f_{\mathcal{C}_{fix}}$, and $f_{\tcal{C}}=f_{\tcal{C}_{fix}}$ for a fixed frame repair operator.
Therefore, with frame repair operator $\hat{\lambda}$, the marginal distribution of the $j$th logical measurement for the fixed circuit $\tcal{C}_{fix}$ matches that of the ideal circuit $\mathcal{C}$.
By Lemma~\ref{lemma:frame_trivial}, different choices of frame configuration give rise to the same measurement distribution for a fixed circuit.
In particular, if we can show the existence of a frame repair operator $\lambda_j$ that satisfies property 1, then it will have the same marginal measurement distribution for the $j$th measurement under $\tcal{C}_{fix}$.
Conditioned on the previous $j-1$ measurement results, this is the same as the marginal measurement distribution for $\tcal{C}$, thereby completing the proof of property 2.

Let us now prove property 1.
By our assumption, the decoding problems of both the first $j-1$ measurement results and the first $j$ measurement results also satisfy the condition that the fault configuration has largest weight of any connected cluster less than $d/t$.
By Lemma~\ref{lemma:trivial_action}, there exists a frame repair operation $\hat{\lambda}|_{j-1}$ and hence frame configuration $\hat{h}|_{j-1}=g\oplus\hat{\lambda}|_{j-1}$, such that $P(f|_{j-1})\oplus P(\hat{h}|_{j-1})$ acts trivially on the first $j-1$ logical measurement results.
Similarly for the $j$th logical measurement, there exists $\hat{\lambda}|_j$ and $\hat{h}|_j=g\oplus\hat{\lambda}|_j$, such that $P(f|_j)\oplus P(\hat{h}|_j)$ acts trivially for the first $j$ logical measurement results.

The actual frame configuration we chose for decoding the first $j-1$ measurements, $h|_{j-1}$, may differ from $\hat{h}|_{j-1}$, as we needed to maintain consistency with previously committed measurements.
This may lead to different measurement outcomes for this specific shot (note that the distribution still remains the same).
For joint decoding of the first $j$ measurements, let us therefore choose the following inferred frame assignment
\begin{align}
\lambda|_j=\hat{h}|_{j-1}\oplus h|_{j-1}\oplus \hat{\lambda}|_j.
\end{align}
With this assignment and by linearity, the action on the codespace is
\begin{align}
&P(f|_j)\oplus P(\hat{h}|_{j-1})\oplus P(h|_{j-1})\oplus P(\hat{h}|_j)\nonumber\\
=&\qty[P(f|_j)\oplus P(\hat{h}|_j)]\oplus\qty[P(\hat{h}|_{j-1})\oplus P(h|_{j-1})].
\end{align}

$P(f|_j)\oplus P(\hat{h}|_j)$ has trivial logical action by construction (the hat is present), so the combined action is identical to that of $P(\hat{h}|_{j-1})\oplus P(h|_{j-1})$.
Similarly, $P(f|_{j-1})\oplus P(\hat{h}|_{j-1})$ has trivial logical action by construction, so on the first $j-1$ measurements, we have that
\begin{align}
&P(\hat{h}|_{j-1})\oplus P(h|_{j-1})\nonumber\\=&[P(f|_{j-1})\oplus P(\hat{h}|_{j-1})]\oplus [P(f|_{j-1})\oplus P(h|_{j-1})]\nonumber\\
=&P(f|_{j-1})\oplus P(h|_{j-1}),
\end{align}
which is exactly the same as the final logical action for the decoding problem of the first $j-1$th measurements.
Therefore, for this choice of $\lambda|_j$, the first $j-1$ measurement results are consistent with the previous round of decoding, proving property 1.
\qed

\subsection{Threshold Theorem}
\label{si_sec:threshold}

Using the preceding characterization of errors, we prove our main result in this section, the existence of a threshold below which logical errors are exponentially suppressed in the code distance.

First, we reproduce a lemma that bounds the number of connected clusters of a given size, a core component of a number of fault-tolerance proofs~\cite{aliferis2007accuracy,bombin2015single,kovalev2013fault,gottesman2013fault,kubica2022single}.
We will make use of the presentation from Ref.~\cite{gottesman2013fault}, specializing to the case where the specific set is a single vertex, as is needed for the main theorem.

\begin{lemma}[Counting lemma on vertices, Lemma 5 of \cite{aliferis2007accuracy}]
\label{lemma:counting}
Consider a specific vertex $\alpha$ in a graph for which every vertex has degree at most $v$.
Let $N_v(s,\alpha)$ be the number of connected sets containing $\alpha$ and a total of $s$ vertices (i.e. $s-1$ vertices beyond $\alpha$).
Then \mbox{$N_v(s,\alpha)\leq(v\e)^{s-1}$}, with $\e$ the usual base of the natural logarithm.
\end{lemma}

Using the counting lemma, we can now complete the proof of our main theorem.

\begin{thm}[Fault tolerance of decoding strategy in Def.~\ref{def:decoding_strategy}]
Consider a transversal realization $\tcal{C}$ of an ideal quantum circuit $\mathcal{C}$ (Def.~\ref{def:error_corrected_circuit}),  subject to the local stochastic noise model with probability $p$, with the circuit involving at most $B$ code blocks of an $[[n,k,d]]$ CSS $(r,c)$-LDPC quantum code family, $T$ layers of operations, with a single SE round following each operation, and $M$ logical measurements.
Then there exists a threshold $p_0$, such that for $p<\frac{121}{144}p_0$, the probability $P_{err}$ of either heralded errors or regular logical errors for the entire circuit, when using the decoding strategy in Def.~\ref{def:decoding_strategy}, is at most $C(p/p_0)^{\frac{d}{2t}}$.
Here, $d$ is the code distance, $t$ is the maximal part size in the transversal partition, $p_0=\frac{1}{(96\e cr)^2}$, $C=\frac{MBT(4n-k)}{4\e cr}$.
\end{thm}

First, let us count the number of possible fault locations under our local stochastic error model.
There are $nB$ physical qubits, each of which experiences at most 3 types of errors $(X,Y,Z)$, leading to $3nBT$ possible data qubit fault locations.
Each logical qubit has $n-k$ independent stabilizer generators that we measure, so there are $(n-k)B$ possible stabilizer measurement errors.
Since there are $T$ layers of operations, the number of fault locations in the circuit is at most
\begin{align}
N_f\leq 3nBT+(n-k)BT=(4n-k)BT.
\end{align}
By definition, this is also the number of vertices in the syndrome adjacency graph $\Xi$.

Next, let us bound the number of neighboring vertices for any vertex in $\Xi$.
By definition, the stabilizer weight is upper bounded by $r\geq 1$, while the number of stabilizers each qubit is involved in is upper bounded by $c\geq 1$.
Therefore, each data qubit error can cause an error on at most $c$ stabilizers.
Since we focus on depth-one transversal operations, each stabilizer is involved in at most four detectors (at most 2 in the past and 2 in the future due to the branching detectors for CNOTs).
Therefore, each data qubit error is connected to at most $4c$ edges.
Each measurement error affects a single stabilizer, which is involved in at most four detectors, so the number of edges it is connected to is also upper bounded by $4c$.
Each detector consists of at most three stabilizers for a depth-one transversal operation, each of which is connected to at most $r$ qubits, where each qubit has at most three types of elementary errors under a depolarizing channel.
Together with the three measurement errors on the stabilizers, each hyperedge is connected to at most $9r+3$ error types.
Putting this together, the number of neighboring vertices for any vertex in $\Xi$ is upper bounded by the constant
\begin{align}
v\leq 4c(9r+3)\leq 48cr.
\end{align}

Suppose a given fault configuration involves $s$ faults.
As we are using the MLE decoder, this implies that the error $e$ must involve at least $\lceil s/2\rceil$ faults.
Since each fault has probability at most $p$, the probability that this fault configuration appears is at most
\begin{align}
\sum_{i=\lceil s/2\rceil}^s {s \choose i}p^i\leq \qty[\sum_{i=0}^s {s \choose i}]p^{s/2}=2^s p^{s/2}.
\end{align}

For each logical measurement, by Lemma~\ref{lemma:logical_error_weight}, the fault configuration must involve a connected cluster of at least $d/t$ faults.
Therefore, applying Lemma~\ref{lemma:counting} to $\Xi$ consisting of $N_f$ vertices, the number of connected clusters $N_s$ of size $s$ is upper bounded by the sum of clusters which contain any given vertex:
\begin{align}
N_s\leq N_f (v\e)^{s-1}.
\end{align}

By Lemma~\ref{lemma:logical_error_weight}, if none of the first $j$ rounds of decoding involve a connected cluster of size at least $d/t$, then the decoding strategy (Def.~\ref{def:decoding_strategy}) will not output FAIL, and the output measurement distribution of the first $j$ measurement results will be the same as the ideal distribution.
Since there are $M$ measurements in total, taking the union bound, the total probability $P_{err}$ of outputting FAIL or having a logical error is at most
\begin{align}
P_{err}&\leq M\sum_{s=d/t}^\infty N_s 2^s p^{s/2}\nonumber\\
&\leq M\sum_{s=d/t}^\infty N_f(v\e)^{s-1}(2\sqrt{p})^s\nonumber\\
&= \frac{MN_f}{v\e}\frac{(2v\e\sqrt{p})^{d/t}}{1-2v\e\sqrt{p}}\nonumber\\
&\leq \frac{MBT(4n-k)}{48\e cr}\frac{(96\e cr\sqrt{p})^{d/t}}{1-96\e cr\sqrt{p}}\nonumber\\
&\leq \frac{MBT(4n-k)}{48\e cr(1-96\e cr\sqrt{p})}\qty(\frac{p}{1/(96\e cr)^2})^{d/2t}
\end{align}

The threshold is thus given by $p_0=1/(96\e cr)^2$, but because the summation still goes to infinity exactly at the threshold, we choose to work below a slightly smaller value than the threshold in order to have a finite constant prefactor.
The prefactor can be tuned if one constrains the range of error rates $p$ differently.
Choosing $p<(11/12)^2 p_0$, we have
\begin{align}
P_{err}\leq C\qty(\frac{p}{p_0})^{d/2t},
\end{align}
where
\begin{align}
p_0&=\frac{1}{(96\e cr)^2},\label{eq:p0}\\
C&=\frac{MBT(4n-k)}{4\e cr}.\label{eq:c}
\end{align}

\qed

This theorem demonstrates that below a certain physical error threshold, the logical error rate is exponentially suppressed in the code distance.
Thus, despite never requiring $d$ rounds of syndrome measurement anywhere in the circuit, we can still maintain fault-tolerance.
As is the case with threshold proofs, many of the bounds here are loose and the actual threshold will be much higher, as we demonstrated numerically.
While we assumed magic state inputs with known stabilizer values for this theorem, we conjecture that the same techniques, when applied jointly to magic state distillation and the main computation, will still yield a $\Theta(d)$ saving when classical decoding is sufficiently fast.

The logical error rate of our protocol scales linearly with the space-time volume of the original circuit.
As only a single SE round follows each operation, another potential benefit of our approach is that there are fewer potential error locations, which may lead to a more favorable constant factor for the logical error rate.
This can partially offset any reduction in the threshold, which our numerics find to be rather minimal.

We can directly apply this theorem to the case of the surface code, plugging in the specific constants.
This leads to the following theorem:

\begin{thm}[Fault tolerance for any ideal quantum circuit with magic state inputs]
Consider a surface code transversal realization $\tcal{C}$ of an ideal quantum circuit $\mathcal{C}$ (Def.~\ref{def:surface_code_circuit}), subject to the local stochastic noise model with probability $p$, with the circuit involving at most $B$ code blocks of a $[[d^2+(d-1)^2,1,d]]$ non-rotated surface code, $T$ layers of operations, with a single SE round following each operation, and $M$ logical measurements.
Then there exists a threshold $p_0$, such that for $p<\frac{121}{144}p_0$, the probability $P_{err}$ of either heralded errors or regular logical errors for the entire circuit, when using the decoding strategy in Def.~\ref{def:decoding_strategy}, is at most $C(p/p_0)^{\frac{d}{4}}$.
Here, $d$ is the code distance,  $p_0=\frac{1}{(1536\e)^2}$, $C=\frac{MBTd^2}{8\e}$.
\end{thm}

For the non-rotated surface code, we have $n=d^2+(d-1)^2\leq 2d^2$, $r=c=4$, $t=2$.
$t=2$ comes from the fold-transversal $S$ gate, leading to a $d/4$ scaling exponent, but this can likely be improved with more careful error analysis.
Plugging this into Eqs.~(\ref{eq:p0},\ref{eq:c}), we have
\begin{align}
p_0&=\frac{1}{(96\e cr)^2}=\frac{1}{(1536\e)^2}\\
C&=\frac{MBT(4n-k)}{4\e cr}\leq \frac{8d^2MBT}{64\e}=\frac{MBTd^2}{8\e}.
\end{align}
\qed

\subsection{Single-Shot Patch Growth}
\label{sec:fault_tolerance_magic_state}

We now extend our results to the case where the magic state input has a smaller code distance $d_1$ that is grown to the full distance $d$ in a single EC round.

\begin{figure}
\centering
\includegraphics{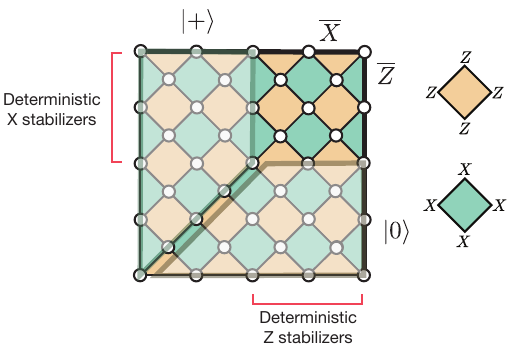}
\caption{Illustration of logical qubit growth process for the surface code.
The initial $d_1=3$ logical qubit, located on the top right, is grown into a larger logical qubit with $d=5$ by initializing the qubits in the top left in $|+\rangle$ and the bottom right in $|0\rangle$.
The strips indicated in red have deterministic stabilizer values, which leads to a lower bound on the weight of an undetectable logical error.}
\label{fig:patch_growth}
\end{figure}

\begin{definition}[Single-shot patch growth]
\label{def:single_shot_patch_growth}
Given an ideal quantum circuit $\mathcal{C}$ with magic state inputs and feed-forward operations (Def.~\ref{def:ideal_circuit}), we define its surface code transversal realization with reduced magic state inputs $\mathcal{C}_m$, with distances $(d,d_1)$, as the surface code transversal realization of distance $d$ defined in Def.~\ref{def:surface_code_circuit}, together with the following operations:
\begin{enumerate}
\item Initialization of some sets of logical qubits of distance $d_1\leq d$ in state $\lket{T}=T\lket{+}$, and with all stabilizer values fixed to +1, up to local stochastic noise on each physical qubit of strength $p$.
\item Logical qubit block growth~\cite{li2015magic,lao2022magic} from distance $d_1$ to $d$, by performing the initialization in the pattern shown in Fig.~\ref{fig:patch_growth} and performing one SE round.
\end{enumerate}
\end{definition}

Although we grow the patch in a single step, the information provided by transversal measurements still allows us to maintain a code distance of $d_1$.
Our discussion focuses on the surface code case, although it is likely that this can be extended to other scenarios.
We use the same decoding strategy defined in Def.~\ref{def:decoding_strategy}.
The statement is then essentially the same as before, except the distance $d$ is replaced by $d_1\leq d$, the size of the magic state input.
This is because all $Z$/$X$ stabilizers in the region highlighted in red in Fig.~\ref{fig:patch_growth} are deterministic, and a chain of errors that spans this region, in order to produce a logical error, must have weight at least $d_1$.
We can readily generalize the rest of our results to prove an analogous fault tolerance theorem, with the distance replaced by the reduced distance of the magic state input.

\section{Limiting Errors in Multi-Stage Distillation Factories}

\begin{figure}
    \centering
    \includegraphics[width=1.0\linewidth]{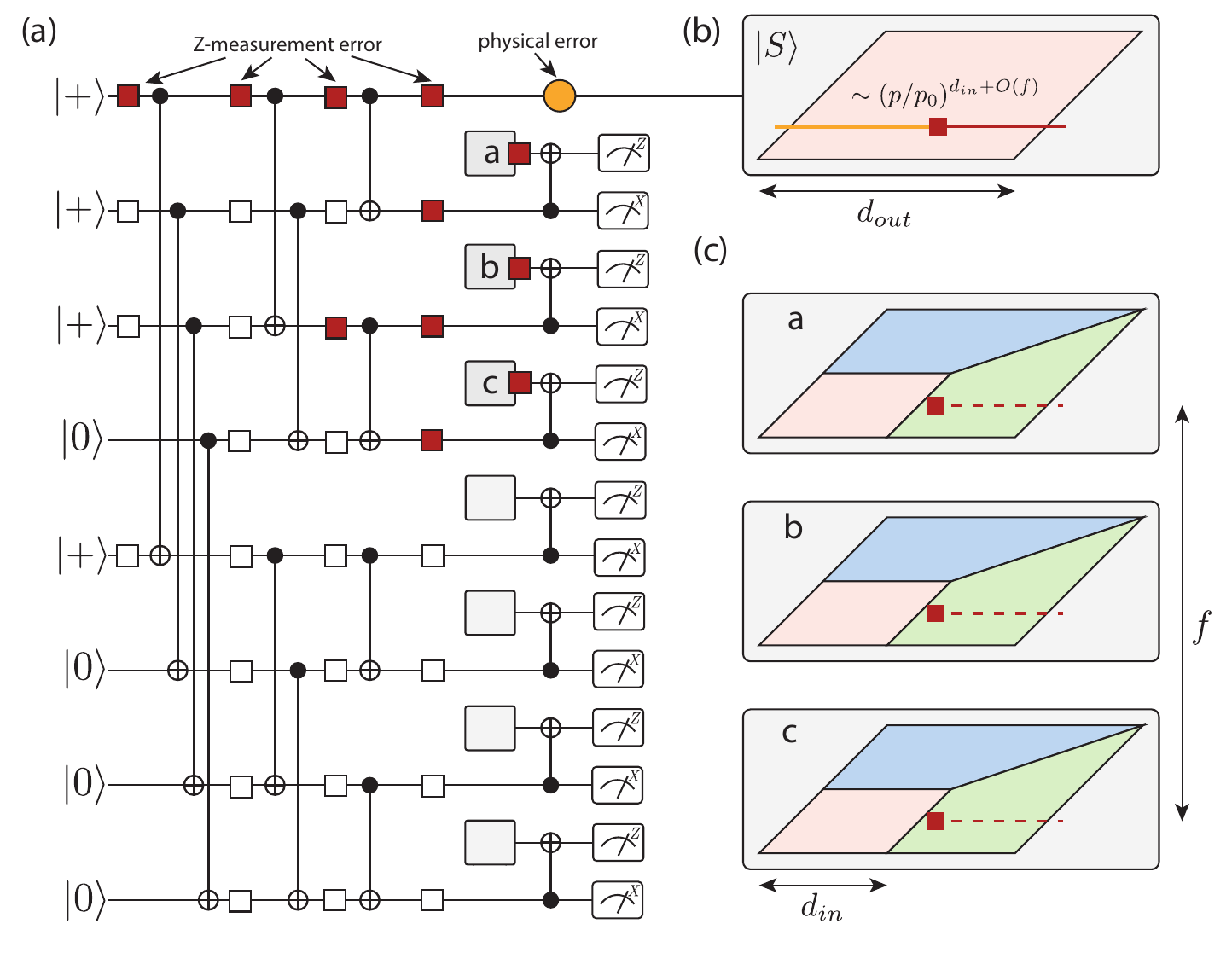}
    \caption{(a) We consider a multi-stage distillation factory, where the inputs (gray boxes) are protected to distance $d_{in}$, and then grown via $O(1)$ rounds of syndrome measurement. A correlated set of measurement errors, starting from the inputs, snaking through the factory, and terminating on the output qubit, can propagate a non-local frame variable to the output qubit. (b) When combined with a physical error of weight $d_{in}$, this creates a genuine logical error. (c) The structure of the measurement errors is associated with a distance $f$ error of the distillation code.}
    \label{fig:low_weight_distillation}
\end{figure}

There is an interesting interplay that occurs when we combine $O(1)$ SE rounds per logical operation with a multi-stage distillation factory. In a multi-stage factory, we input a target state (i.e. $\lket{S}$ or $\lket{T}$) prepared at some small distance $d_{in}$, and use single-shot patch growth to inject the state into a larger code with distance $d_{out}$.
The distillation code, based on a distance $f$ code with a suitable transversal gate, is then used to improve the quality of the magic state.
In the absence of measurement errors, the resulting logical state should be protected to distance $\min\{d_{in} f,d_{out}\}$.
However, we find that performing $O(1)$ SE rounds per logical operation causes the error protection to go from multiplicative ($d_{in} f$) to additive ($d_{in}+f$), due to error correlations between multiple patches.
Therefore, care must be taken when designing multi-stage factories based on transversal gates to ensure that the output achieves the target distance.

We illustrate the structure of such errors in Fig.~\ref{fig:low_weight_distillation}.
Consider the set of $Z$ stabilizer measurement errors illustrated in Fig.~\ref{fig:low_weight_distillation}(a), where the errors occur at the same transversal coordinate of the code patch.
This set of $O(f)$ measurement errors is undetectable by the factory, but combined with the frame variable, it generates a physical error-string of weight $d_{out} - d_{in}$ on the output.
As shown in Fig.~\ref{fig:low_weight_distillation}(b), when combined with a physical error chain of length $d_{in}$, it produces a genuine logical error on the output state, which cannot be detected and corrected by future transversal measurements.
Thus, the output state in this multi-stage factory is only protected to distance $O(d_{in} + f)$.

This example highlights the care needed when analyzing state injection in the transversal setting with $O(1)$ SE rounds per logical gate.
Our numerics in Fig.~4 of the main text shows that these issues are alleviated at small code distances and moderate logical error rates, where we do not see any deviation from the leading-order multiplicative behavior.
Practical magic state distillation factories often employ $d_{in}\approx 15$ and $d_{out\approx 30}$, so it is likely that inserting a few more SE rounds during state injection, comparable to the depth of the Clifford part of the distillation factory, will be sufficient to suppress these correlated errors to a sufficiently low level.

\section{Necessity of Restoring Consistency}
\label{si_sec:consistency}
In this section, we provide an explicit example of how the usual decoding procedure has a high chance (linear in the physical error rate $p$) of changing the assignment of a logical measurement result, and how the step of restoring consistency addresses this issue.

\begin{figure}
\centering
\includegraphics[width=\linewidth]{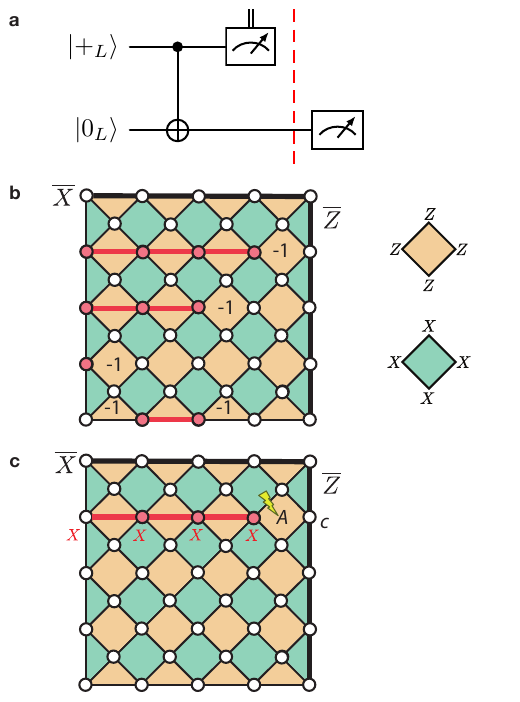}
\caption{(a) Logical circuit illustrating how future syndrome information changes the interpretation of logical measurement results with probability linear in the physical error rate. The red dashed line separates the two rounds of decoding, and there is an $O(p)$ chance that the assignments disagree.
(b) Illustration of the rule for pairing the initial random stabilizer values, where we pair all excitations to the left boundary. Note that unlike the usual decoding problem, all patterns of Z stabilizer values are equally likely, so any pairing rule works.
(c) Illustration of how an incorrect inference of a single stabilizer (A) will lead to a string of $X$ errors.}
\label{fig:consistency_example}
\end{figure}

As shown in Fig.~\ref{fig:consistency_example}(a), we consider two logical qubits, one prepared in $\lket{+}$ and one in $\lket{0}$, followed by a transversal CNOT. Next, we measure the first logical qubit in the $Z$ basis and commit to its measurement result with the existing information, as it may be used for feed-forward operations on other logical qubits. Finally, we measure the second logical qubit in the Z basis, and perform joint decoding across the system. Each logical qubit is a surface code, and for pairing the initially random Z stabilizers when initializing $\lket{+}$, we choose a convention that always pairs to the left boundary, as illustrated Fig.~\ref{fig:consistency_example}(b). Since the initial projection of stabilizers is fully random, any convention here is valid.

We assume the following simplified error model to illustrate that there is an $O(p)$ chance that the two decoding results disagree: each syndrome measurement has error $p$, the final transversal measurement has error $p+\epsilon$, and there are no other errors.

Now, consider a single stabilizer measurement error on the $Z$ plaquette labeled by $A$ in the top logical qubit $\lket{+}$, as illustrated in Fig.~\ref{fig:consistency_example}(c).
When performing the first round of decoding, in which we only have syndrome information up to the first logical qubit measurement, we will observe an inconsistency between the stabilizer value of $A$ measured in the first round, and that inferred from the transversal logical measurement.
This has two possible explanations of low weight:
\begin{enumerate}
\item The stabilizer $A$ was measured incorrectly during initialization, with probability $p$.
\item The data qubit $c$ was measured incorrectly during readout, with probability $p+\epsilon$.
\end{enumerate}

Since the latter event has higher probability, we will choose correction (2). Together with the $X$ string chosen during initialization, this causes a logical $X$ flip on the readout value.

Now consider the second round of decoding, which happens when the second logical qubit is also measured. Upon performing correlated decoding, the additional syndrome information on the bottom patch is consistent with $A$ being measured incorrectly and the data qubit being measured correctly. Therefore, the probability of case (2) above is now updated to $(p+\epsilon)^2<p$, so we choose correction (1). This leads to no logical $X$ flip on the readout value.

Comparing the two cases, we see that the first round of decoding in the presence of partial information resulted in a logical flip, while the second round of decoding, due to its extra syndrome information, did not result in a logical flip. In other words, the logical measurement results differ.

This shows that the probability that the decoding result changes when more syndrome information is available scales linearly with the physical error rate, rather than $p^{(d+1)/2}$, and therefore a direct application of standard procedures does not achieve fault tolerance in the presence of non-FT state prep and feed-forward operations.

Our procedure addresses this issue by using the application of a logical Pauli stabilizer to restore consistency:
\begin{enumerate}
\item When decoding the first logical qubit LQ1, our assignment of errors incorrectly applies a logical $X$ flip to LQ1, relative to the value that an oracle with access to the actual errors will choose. However, since the marginal logical outcome was 50/50, the distribution is unchanged by this flip.
\item When decoding the second logical qubit, we conclude that a logical $X$ flip should not have been applied to LQ1. However, since this is inconsistent with our previous commitment, we apply the logical stabilizer $X_1X_2$, which flips both LQ1 and LQ2. This guarantees that we assign the same result in both steps to LQ1.
\item Because the marginal distribution of the first step and the conditional distribution in the second step are correct with high probability, and we assign the same result to the same qubits in different steps, the joint logical measurement distribution is correct with high probability.
\end{enumerate}

\section{Example: Repetition Code}

We now consider a simple illustrative example of the fault-tolerance approach.
For illustration purposes, we will focus on a repetition code example~\cite{haah2024what}, although the lessons readily generalize to the surface code.

\begin{figure}
\centering
\includegraphics[width=\columnwidth]{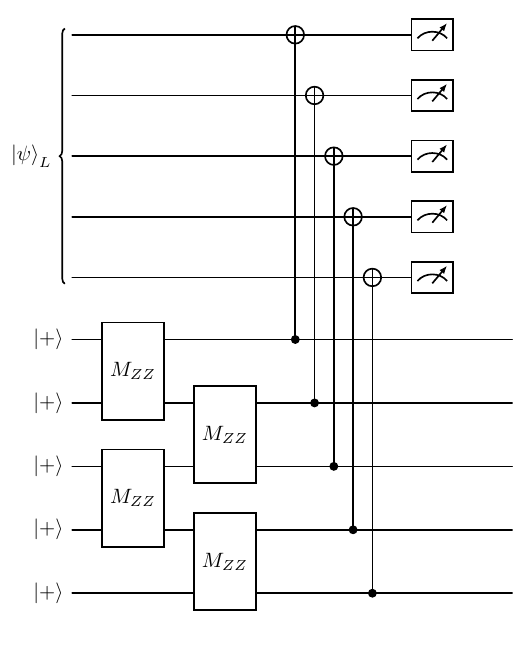}
\caption{Repetition code example. The bottom logical qubit is prepared via a single round of stabilizer measurements, and then executes a transversal CNOT on the top logical qubit $|\psi\rangle_L$.
Although the state preparation of the bottom logical qubit is not fault-tolerant in the conventional sense, we are still able to reproduce the logical measurement statistics of an ideal circuit with high probability as the distance $d\rightarrow \infty$.}
\label{fig:repetition_code_example}
\end{figure}

Consider two repetition code logical qubits.
The first is prepared in an unknown logical quantum state $\lket{\psi}$, while the second logical qubit is prepared in a single-shot manner in the $\lket{+}$ state, by preparing all physical qubits in $|+\rangle$ and measuring neighboring $ZZ$ stabilizers once.
At this stage, the physical state of the second logical qubit is in fact a mixture of product states~\cite{hastings2011topological} if we try to directly fix the stabilizer values back to the code space.
This is because we have not gained sufficient confidence about the $ZZ$ stabilizer values from the single faulty measurement, and therefore may incorrectly pair up excitations, potentially causing a larger string of $X$ errors.
Importantly, however, it is not yet necessary to fix the stabilizer values back to the code space, as we have not yet performed a logical measurement.
The transversal logical measurement later on will help us avoid harmful long $X$ error strings due to its reliable syndrome information.

We then perform a transversal CNOT, with the second logical qubit as control and first logical qubit as target.
Following this, we measure the first logical qubit in the $Z$ basis.
With Pauli feedforward, this circuit can teleport the unknown state $\lket{\psi}$ to the second logical qubit.

At this point, we may naively be concerned about the correctness of the first measurement result, since the string of $X$ errors can lead to a probability linear in the physical error rate $p$ of flipping this measurement result.
However, the situation is a bit more subtle: when defining correctness of a quantum computer execution in a model of classical inputs and outputs, what we really care about is that the ideal measurement distribution is reproduced, rather than a given shot being interpreted in a particular way.
In the circuit that we are executing here, the logical CNOT propagates the randomness to the first logical qubit, and thus the measurement result will be a 50/50 random number.
By itself, flipping the logical readout result therefore does not change the measurement distribution of the first logical qubit measurement, and so in a sense, the long $X$ error string does not yet cause a logical error at this stage.
More broadly, consider any logical measurement where we may be worried about large error strings from some Pauli initialization.
In order for the error string to have an effect, the initialized Pauli stabilizer, upon propagating through the Clifford circuit, must anti-commute with the logical measurement.
This, however, implies that the measurement result was random, so a flip of the measurement result does not change the measurement distribution.

Although the measurement distribution for the first logical qubit is unchanged, this does not yet mean the whole circuit is executed correctly: we still need to guarantee that the joint distribution between all logical measurements is the same as the ideal circuit.
To understand this, we need to provide more specification on our unknown state $\lket{\psi}$.
In particular, we need to specify whether we have already prepared it in a fault-tolerant fashion, such that the residual noise on it is local stochastic, or whether some of the stabilizers have not yet been fault-tolerantly assigned.

First, consider the former case, where we have already fault-tolerantly prepared the unknown magic state $\lket{\psi}$ through some method.
For the surface code, this may come from another circuit that involved e.g. magic state distillation.
The transversal measurement of the first logical qubit reveals information about the product of stabilizers at the same location on the two logical qubits, up to local stochastic errors, since we directly measure the physical qubits and therefore errors can be regarded as data errors rather than syndrome errors.
Since we know the stabilizers of the first logical qubit with only local stochastic errors, we have also effectively made inferences about the stabilizer initialization values of the second logical qubit.

In the second case where the first logical qubit also has unknown stabilizer initialization values, its preparation must trace back to some Pauli basis input state.
For example, consider the case where the first logical qubit was also initialized in a single step in $\lket{+}$.
The transversal logical measurement still reveals information about the product of stabilizers, but now we no longer learn the initialization values of each of the stabilizers.
Fortunately, this is not a concern, as only the product of stabilizers is relevant to interpreting the logical measurement result.
Later logical measurements will give us additional information that will allow us to learn the individual values of stabilizers when they are necessary.
Indeed, we can extend the intuition of anti-commutation between logical measurements and logical Pauli stabilizers discussed above.
Products of multiple logical measurements that anti-commute with some Pauli initialization are random, and therefore their distribution does not get affected by logical flips.
Products that commute with all Pauli initializations, on the other hand, are insensitive to the large error string due to the commutation, and therefore are not affected either.
Because Pauli initializations propagate to Pauli products through the Clifford circuit, and all logical measurements are in the Pauli basis, the two must either commute or anti-commute.

When we now measure the second logical qubit, in our decoding strategy, we will re-decode the existing portion of our circuit.
This may cause a different assignment of the first logical measurement result.
However, we can apply an $\overline{X}$ operation at initialization on the second logical qubit, which doesn't change the $\lket{+}$ state.
Propagating this $\overline{X}$ flip through, this will flip both logical measurement results, flipping the first measurement back to being consistent with the previous measurement, while also flipping the second measurement result.
We thus interpret the second measurement result as having taken the flipped value, so that we maintain consistency with the first measurement.
With this method, our theorem shows that the measurement distribution of the noisy circuit can be made arbitrarily close to the ideal circuit, as the code distance is increased.

\section{Example: Non-Clifford Operations}
\label{si:non_clifford}

In this section, we discuss the example in Fig.~2 of main text in more detail, where we perform $\lket{T}$ state teleportation and feed-forward operations.

Again, one might be worried about making an incorrect commitment to the measurement result used for teleportation, since a non-trivial feed-forward $S$ gate has been applied (Fig.~2(b) of main text).
However, as illustrated in Fig.~2(c) and discussed throughout our paper, applying an $\overline{X}$ on the $\lket{+}$ initial state does not change the state.
Propagating this through, we find that the combination of an $X$ operator on the bottom qubit and a $Y$ on the middle qubit also stabilizes the state.
Thus, if we infer a different logical measurement result for the bottom qubit later on, we can flip it back to our originally-committed result, as long as we also apply a $Y$ to the middle qubit.

Another possible concern is how a logical measurement result that, due to a magic state input, is no longer deterministic or 50/50 random would be affected by the non-fault-tolerant Pauli basis initialization.
However, as discussed in the previous section, a logical measurement that can be affected by the large error string originating from a Pauli basis initialization must by necessity, also anti-commute with the initial logical stabilizer, ensuring that it will be a 50/50 random variable.
This is indeed the case for the circuit illustrated in Fig.~2 of the main text.
Otherwise, the relevant basis has deterministic stabilizers to begin with on all input logical qubits, and errors can be appropriately detected and corrected.

\section{Analysis of Single-Round Lattice Surgery}
In this section, we analyze single-round lattice surgery in more detail, and explain why unlike the transversal case, it is not fault-tolerant.
We note that our example is very similar to the one discussed in Appendix D of Ref.~\cite{kim2022fault}.
Our analysis indicates that the scheme proposed there is not fault-tolerant, although suitable modifications based on transversal algorithmic fault tolerance should be able to recover most of their conclusions.

We analyze a variant of the circuit shown in Fig.~3 of the main text.
Here, instead of preparing the bottom three qubits in $\lket{0}$, we prepare them in some arbitrary quantum state $\lket{\psi}$, with known stabilizer values up to local stochastic noise.
This closely mirrors the typical situation in a deep circuit.
We perform a transversal CNOT from the GHZ state to three qubits initialized in $\lket{\psi}$, and then measure the original GHZ qubits in the $\overline{X}$ basis.
With a $\overline{Z}$ feed-forward on each qubit, the correlations of the GHZ state are now imprinted onto the bottom 3 qubits.
However, with state preparation based on lattice surgery, knowledge of the specific GHZ state we prepare relies on obtaining the product of values of $Z$ stabilizers of the larger surface code patch, along the seams between the different logical qubits.
More specifically, labeling the logical qubits with 1 to 3 from top to bottom, the correlator $\overline{Z}_1\overline{Z}_2$ is initialized to a random value when we perform the initial random projection of the larger surface code.
In the absence of errors, this correlator will be equal to the product of $Z$ stabilizers along the corresponding boundary.
However, a single measurement error can cause us to misinterpret $\overline{Z}_1\overline{Z}_2$, and we have no way of obtaining and correcting this error later on.
Therefore, in the case of single-shot lattice surgery, a single physical error can lead to a logical error.
Note that we measure the GHZ state in the $\overline{X}$ basis, so that it is not possible to deduce $\overline{Z}_1\overline{Z}_2$ directly through the logical measurements.

We can contrast this with our transversal algorithmic fault tolerance construction.
In this case, even if later decoding steps assign a different logical measurement result to the ancilla qubit, we can apply a frame logical variable to obtain the same result as our previous commitment.
The transversal measurement also ensures that no harmful error events can terminate on the measurement time boundary, and therefore there are no time-like errors that flip the logical measurement result, as occurs in the case of lattice surgery.
Thus, single-shot logical operations are fault-tolerant in the transversal scheme, but not in the case of lattice surgery.

A key distinction between our transversal construction and lattice surgery is thus how the logical information is measured.
For transversal gates, we always directly access the logical information through transversal measurements, in the process obtaining the relevant information to process and correct errors and interpret logical measurements correctly.
In contrast, the logical information is contained in noisy syndrome measurements for lattice surgery, thereby necessitating repetition before one can gain confidence about the results.

\section{Details of Numerical Simulations}
\label{si:numerics}

Here we describe the numerical simulations conducted to evaluate the performance of our decoding strategy. 
To simulate a logical circuit, we first generate a description of the physical circuit and noise model using Stim~\cite{gidney2021stim}, an open-source Clifford simulation package. 
From this description, we specify the detectors and logical observables of the circuit.
Because in practice Stim requires logical observables to be deterministic under noiseless exectution, we label non-deterministic logical observables as \textit{gauge detectors}, whose ideal measurement outcome can be non-deterministic. 
We then use Stim to Monte-Carlo sample the detectors and logical observables over different physical noise realizations.
Each sample is decoded using our decoding strategy, with each logical measurement interpreted using only the partial syndrome information up to that point, and a logical error is observed if either a heralded inconsistency or a regular logical error occured.
The logical error rate for a given circuit is computed from the mean over many Monte-Carlo samples, and the error bars correspond to the Clopper-Pearson confidence interval based on a Beta distribution with a significance level of $0.05$.

We specify the physical operations used to generate the rotated surface code logical operations following Definition~\ref{def:surface_code_circuit}.
In addition to these operations, we also allow physical measurements and initialization in the $X$ basis (rather than using a $H$ operation plus measurement or initialization in the $Z$ basis). 
We perform a SE round by using a sequence of four physical CNOTs to map each stabilizer value to an ancilla qubit, using the gate ordering described in Ref.~\cite{acharya2023suppressing}.
Because our main result enables $O(1)$ rounds of SE between transversal CNOTs, we have flexibility in where SE is performed within the circuit.
In Figs.~3(a-b) and Fig.~4(c-d), for example, we perform one SE round after each transversal CNOT on the logical qubits involved in the gate.
In contrast, no intermediate SE rounds are performed in Fig.~3(d).

We add noise to each physical operation using a  circuit-level noise model similar to  Ref.~\cite{higgott2023improved}. 
Concretely, for a chosen physical error rate $p$, we add a depolarizing channel with probability $p$ to each physical operation. We apply a two-qubit depolarizing channel after each entangling gate, a single-qubit depolarizing channel after each single-qubit gate and initialization, and a single-qubit depolarizing channel before each measurement. 
In contrast to Ref.~\cite{higgott2023improved}, we do not apply noise to idling qubits during measurement and initialization. 
However, we do apply a single-qubit depolarizing channel to idling qubits during gate operations.

For the $\lket{Y}$ state distillation factory simulations in Fig.~4 of the main text, we perform state injection from the corner qubit at distance $d_0=3$ (Fig.~4b), following the procedure described in Ref.~\cite{lao2022magic}.
More precisely, we perform two rounds of SE during the first phase of state injection, growing the patch size from one to $d_0$, and post-select on having consistent stabilizer values between the two rounds as well as the correct stabilizer value for the deterministically-initialized stabilizers. Then we perform the single step patch growth from distance $d_0$ to $d_1$.
In order to probe the performance of the state distillation factory without prohibitive sampling costs, we add extra $Z$ errors with probability $p_{Z}$ on the injected physical qubit to increase the error rate.
The output state infidelity is probed by performing a noiseless $\overline{S}$ rotation via $S$ gate teleportation with a noiseless-injected ancilla patch and performing an $X$ basis measurement.

For the single-level distillation factory simulations, we set $d_0 = 3$ and vary $d_1$ in the set $\{3, 5, 7, 9\}$. Each data point in Fig. 4(c) represents $10^5$ samples after post-selection during the state injection step. After generating a sample of measurement results that succeeded in all state injection checks, we first partially decode the logical $Z$ basis measurements on all injected qubits and logical $X$ basis measurements on the remaining qubits other than the output qubit to filter out factory failures. Then, we decode all qubits to estimate the output infidelity of the $\lket{Y}$ state.
For the two-level distillation factory simulation, the input states of the second-level factory are the output states from the first-level factories followed by a single step patch growth from distance $d_1$ to $d_2$. We set $d_0 = 3$, $d_1=5$, and $d_2=9$.
To decode the two-level factory efficiently in practice, we first sample measurement results for the entire circuit.
We then decode each level-1 factory and discard runs in which any of the physical state injections or level-1 factories failed, in order to reduce the computational cost.
For instances where all level-1 factories succeeded, we perform correlated decoding on the entire level-2 factory, with the output $\lket{Y}$ state rotation and measurement, to determine the output and estimate the logical error rate.
Assuming $p_Z = 10\%$ and $p = 0.1\%$, we obtain $99620000$ raw shots in total, of which $312825$ shots passed all factory checks, and $3$ logical errors were observed.

As the state injection protocol itself is noisy, the infidelity of the injected logical state $p_{\mathrm{inj}}$ is greater than $p_Z$. In order to estimate the infidelity of the input logical $\lket{Y}$ state, we simulate the state injection protocol itself by injecting a $\lket{Y}$ state, followed by a perfect $\overline{S}$ gate and an $X$ basis measurement described above. All input infidelities in Fig.~4(c-d) refer to $p_{\mathrm{inj}}$ instead of $p_Z$.

Finally, our main Theorem assumes that partial decoding is performed using the MLE decoder, which in practice may have a runtime that grows exponentially with the size of the decoding problem.
However, in practice we find that our decoding strategy still yields a threshold with belief propagation augmented hypergraph union find (belief-HUF), an efficient decoder which runs in polynomial time~\cite{delfosse2022toward, cain2024correlated}. The detailed implementation of belief-HUF for partial decoding is described as follows. We generate the partial decoding hypergraph $\Gamma|_j$ as well as its decomposed version $\Gamma'|_j$ at each partial decoding step $j$. 
Here, $\Gamma'|_j$, which contains essential hyperedges only, can be generated by setting $\mathtt{decompose\_errors = True}$ in Stim from $\Gamma|_j$~\cite{gidney2021stim}. 
Given a sampled detector configuration, we first perform $\mathtt{bp\_{rounds}=5}$ rounds of belief propagation to update the posterior probabilities of error mechanisms for $\Gamma|_j$ and transfer these probabilities into $\Gamma'|_j$. Finally, we apply a hypergraph union-find decoder on $\Gamma'|_j$ with the hyperparameter $\mathtt{weight\_exponent} = 0$ to obtain the decoded logical observables~\cite{cain2024correlated}.

\begin{figure}
    \centering
    \includegraphics[width=\columnwidth]{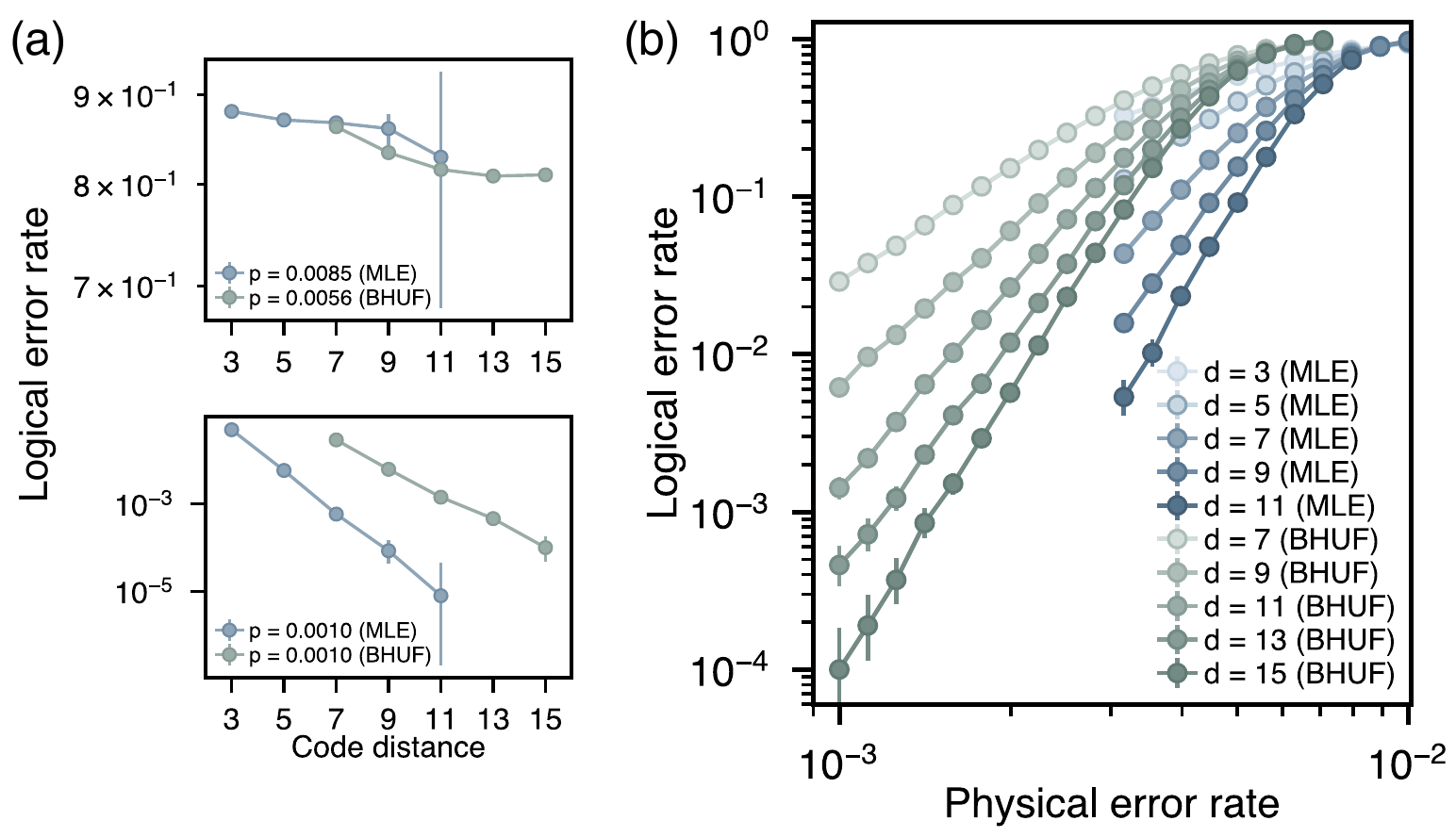}
    \caption{Numerical results for decoding repeated Bell pair measurements with the belief-HUF and MLE decoders. (a) The total logical error rate decreases with the code distance at $p=0.56\%$ for belief-HUF and at $0.85\%$ for MLE (top); the same trend at $p=0.1\%$ (bottom). (b) The total logical error rate as a function of the physical error rate for belief-HUF and MLE.}
    \label{fig:bhuf}
\end{figure}

Fig.~3a in the main text presents the total logical error rate $P_L$, which is the probability of either a heralded error or a regular logical error occurring, as a function of the physical error rate $p$, using the MLE decoder. In Fig.~\ref{fig:bhuf}(b), we show the corresponding results for belief-HUF as well.
These simulations imply the presence of a threshold when using the belief-HUF decoder in practice. As the total logical error rates approach their upper bounds at physical error rates near the thresholds, we cannot precisely estimate the threshold by fitting the universal scaling hypothesis. Nevertheless, we can still estimate a lower bound of the belief-HUF and MLE thresholds by identifying the highest physical error rate at which $P_L$ monotonically decreases as $d$ increases, as shown in Fig.~\ref{fig:bhuf}(a).
We estimate that the threshold for the MLE decoder is $\gtrsim 0.85\%$ and for the belief-HUF decoder is $\gtrsim 0.56\%$, consistent with previous simulation results~\cite{cain2024correlated}.
We expect future optimizations of the decoder to further improve the performance and bring it closer to the MLE decoder.

\section{Comparison with Existing Approaches to Single-Shot Quantum Error Correction}

In this section, we contrast our approach with existing approaches to single-shot quantum error correction.
We highlight the crucial distinction between single-shot quantum error correction, which analyzes an error correction gadget individually, and single-shot logical operations, which applies also to logical operations and analyzes fault tolerance as a whole.

The concept of single-shot quantum error correction was originally proposed by Bombin~\cite{bombin2015single}.
Here, redundancies are present in the syndrome extraction results, allowing one to robustly infer the actual stabilizer values up to small residual errors, in a fashion similar to classical error correction on the syndrome readings.
These ideas were later extended to certain families of quantum low-density parity-check (qLDPC) codes~\cite{quintavalle2020single,campbell2019theory,fawzi2018constant,gu2024single}, where expansion and the so-called confinement property lead to single-shot QEC for quantum memories.
In this case, however, there are usually no stabilizer redundancies, and so the randomly initialized stabilizer values cannot be reliably inferred in the conventional FT strategies.
Here, one only guarantees that the output error after a round of error correction is controlled if both the input error and added noise are controlled, and one may still require $d$ rounds of repetition to learn the initialized values of the stabilizers with sufficient confidence for the individual state preparation gadget.

When considering a full-fledged FTQC, the time cost may be modified, and logical operations are often no longer single-shot.
As mentioned above, state initialization for LDPC codes using conventional FT constructions may require $d$ rounds of repetition, as the values of randomly initialized stabilizers need to be learned reliably.
Moreover, the most general methods for performing logical operations on LDPC codes make use of lattice surgery, which also requires $d$ rounds of syndrome extraction to maintain FT~\cite{cohen2022low,xu2024constant}, similar to the lattice surgery example for the surface code we analyzed.
Therefore, logical gates typically require order of $d$ time cost.
The same consideration also applies to other constant-space-overhead schemes, such as those based on code concatenation~\cite{yamasaki2024time}.
Many logical operations can be implemented in 3D codes in a single-shot fashion~\cite{bombin2015single}, or alternatively using 2D codes and the time direction in a just-in-time decoding fashion~\cite{bombin2018,scruby2022numerical}, but the space-time volume scales as $d^3$, effectively corresponding to a space-time trade-off when compared to the conventional surface code scheme and not leading to a clear advantage~\cite{beverland2021cost}.
As such, while there are multiple approaches with potential promise to produce lower space-time overhead when implementing a generic quantum circuit, to the best of our knowledge, further research is required to show an end-to-end space-time overhead reduction when compared to the standard surface code schemes based on lattice surgery.

In conclusion, single-shot QEC focuses on the fault-tolerance and error-reducing effect of individual error correction gadgets, rather than the complete end-to-end algorithmic context.
This is in contrast to our FT strategy, which uses all accessible information throughout the algorithm, and analyzes the fault tolerance of logical operations.
Our scheme thus has much more forgiving requirements on the code properties, and can serve as a drop-in replacement to existing compilation schemes with an immediate space-time overhead reduction.

\bibliography{si}